\documentclass[12pt]{article}
\usepackage{axodraw,epsfig,a4,amssymb,amsmath}
\newcommand{\pslash}{\not{\hspace{-0.08cm}p}}
\newcommand{\xslash}{\not{\hspace{-0.08cm}x}}
\newcommand{\ggcf}{\frac{g^2 C_F}{16 \pi^2} }

\newcommand{\be}{\begin{equation}}
\newcommand{\ee}{\end{equation}}

\newcommand{\bea}{\begin{eqnarray}}
\newcommand{\eea}{\end{eqnarray}}
\newcommand{\MS}{{\overline{MS}}}
\def\bbbone{{\mathchoice {\rm 1\mskip-4mu l} {\rm 1\mskip-4mu l}
{\rm 1\mskip-4.5mu l} {\rm 1\mskip-5mu l}}}
\title{
\vspace{-2.5cm}
\flushleft{\normalsize DESY 04-021} \\
\vspace{-0.35cm}
{\normalsize Edinburgh 2004/03} \\
\vspace{-0.35cm}
{\normalsize Leipzig LU-ITP 2004/007} \\
\vspace{-0.35cm}
{\normalsize Liverpool LTH 618} \\
\vspace{0.35cm}
\centering{\Large \bf One-loop renormalisation of quark bilinears for
overlap fermions with improved gauge actions}\vspace*{-0.35cm}}
\author{\large R.~Horsley$^1$, H.~Perlt$^{2,3}$, P.E.L.~Rakow$^4$,
G.~Schierholz$^{5,6}$ and A.~Schiller$^3$ \\[0.75em]
-- QCDSF Collaboration -- \\[0.75em]
\normalsize
$^1$ School of Physics, University of Edinburgh, \\
\normalsize 
Edinburgh EH9 3JZ, UK \\
\normalsize
$^2$ Institut f\"ur Theoretische Physik, Universit\"at Regensburg,\\ 
\normalsize
D-93040 Regensburg, Germany \\
\normalsize
$^3$ Institut f\"ur Theoretische Physik, Universit\"at Leipzig, \\ 
\normalsize
D-04109 Leipzig, Germany \\
\normalsize
$^4$ Theoretical Physics Division, Department of Mathematical Sciences,\\ 
\normalsize
University of Liverpool,\\
\normalsize
Liverpool L69 3BX, UK \\
\normalsize
$^5$ John von Neumann-Institut f\"ur Computing NIC,\\
\normalsize
Deutsches Elektronen-Synchrotron DESY,\\
\normalsize
D-15738 Zeuthen, Germany\\
\normalsize
$^6$ Deutsches Elektronen-Synchrotron DESY, \\
\normalsize
D-22603 Hamburg, Germany
}

\date{ }

\begin{document}

\maketitle

\begin{abstract}
We compute lattice renormalisation constants of local bilinear quark operators
for overlap fermions and improved gauge
actions. Among the actions we consider are the Symanzik, L\"uscher-Weisz, 
Iwasaki and DBW2 gauge actions. The results are given for a variety of $\rho$
parameters. We show how to apply mean field (tadpole) improvement to overlap
fermions. 
The question, what is a good gauge action, is discussed from the perturbative
point of view. Finally, we show analytically that the gauge dependent 
part of the self-energy and the amputated Green functions are independent of
the lattice fermion representation, using either Wilson or overlap fermions.
\end{abstract}

\section{Introduction}

Lattice calculations at small quark masses require an action with good chiral
properties. The same is true for calculations of matrix elements of certain
operators, which otherwise mix with operators of opposite chirality. 
Ginsparg-Wilson fermions~\cite{Ginsparg:1981bj} have an exact chiral symmetry
on the lattice~\cite{Luscher:1998pq}, and thus are well suited for these
tasks. A further advantage is that they are automatically $O(a)$
improved~\cite{Capitani:1999uz}. Overlap fermions~\cite{overlap,n2,n} provide 
a four-dimensional realisation of Ginsparg-Wilson fermions. The massless
Neuberger-Dirac operator is defined by
\be
D_N = \frac{\rho}{a} \left( 1+ \frac{X}{\sqrt{X^\dagger X}}\right)\,,\;
X = D_W - \frac{\rho}{a}\,,
\label{over}
\ee
where $D_W$ is the Wilson-Dirac operator, and $\rho$ is a real parameter
 corresponding to a negative mass term. At tree level $\rho$ must
 lie in the range $0 < \rho < 2r$, $r$ being the Wilson parameter.
 Exact chiral symmetry on the lattice has its price however. Numerical
implementations of the overlap operator are significantly more expensive than
Wilson fermions. In spite of this difficulty, calculations with overlap
fermions are progressing rapidly, and we expect to see many more results in
the near future.  

It is important to use a good gauge field action. The cost of the
overlap operator is governed largely by the condition number of $X^\dagger X$.
This number is greatly reduced for improved gauge field
actions~\cite{DeGrand:2002vu}. For example, for the tadpole improved
L\"uscher-Weisz action we found a reduction factor of $\gtrsim 3$ compared to
the Wilson gauge field action~\cite{Galletly:2003vf}. The reason is that the
L\"uscher-Weisz action, as many other improved actions~\cite{Gockeler:1989qg},
suppresses unphysical zero modes, sometimes called dislocations. A reduction
in the number of small modes of $X^\dagger X$ appears also to result in an
improvement of the locality of the overlap operator~\cite{DeGrand:2002vu}.

To obtain continuum results from lattice calculations of hadron matrix
elements, the underlying operators have to be renormalised. A perturbative
calculation of the corresponding renormalisation constants is always the first
step. Recently, the renormalisation constants of
bilinear~\cite{Alexandrou:2000kj,Capitani:2000wi,Capitani:2000aq}  
and four-quark~\cite{Capitani:2000da} operators have been computed to
one-loop order for overlap fermions and Wilson gauge field action. Our aim is
to extend the calculation to general, improved gauge field actions with up to
six links, including the tree-level Symanzik action, the L\"uscher-Weisz
action, the Iwasaki action, and the DBW2 action. In this paper we shall
consider local bilinear quark operators first. Operators including covariant 
derivatives will be treated in a separate publication. Preliminary results of
the calculation have been reported in~\cite{Galletly:2003vf}.

The paper is organised as follows. In Section 2 and Appendix A we give 
the details of the calculation. The basic results are given in Section 3. In
Appendix B we show that the gauge dependent terms in the Green functions
are not only independent of the gauge field action, but also of the type of
fermion. Tadpole improvement is an important issue in lattice perturbation
theory. In Section 4 and Appendix C we tadpole improve our results. Finally,
in Section 5 we give our conclusions. 

\section{Calculational details} 
\label{sec:2}

\begin{figure}[!t]
\begin{center}
\epsfig{file=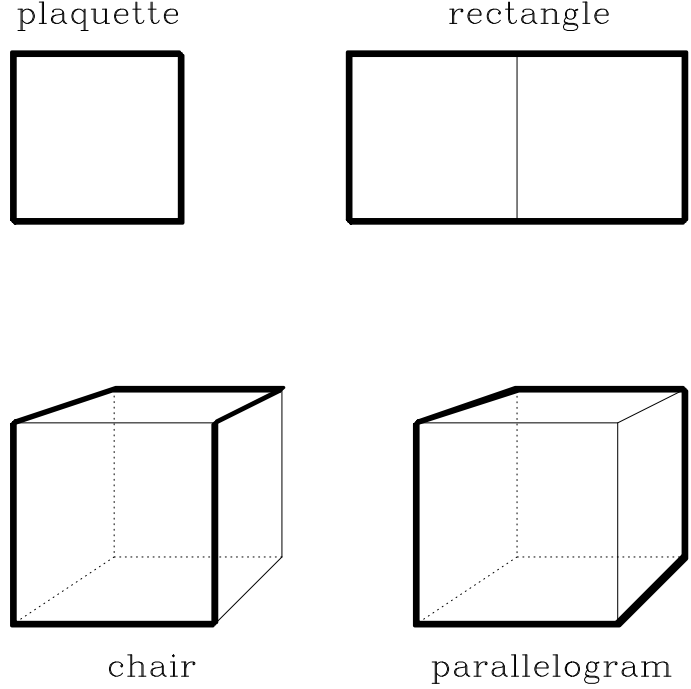,angle=270,width=9cm}
\end{center}
\caption{The meaning of the four terms in the gauge field action
  (\ref{gluonaction}).} 
  \label{cube}
\end{figure}

We denote the lattice action by
\be
S = S_G + S_F \,,
\ee
where $S_G$ is the gauge field action, and $S_F$ is the fermion action. 
Let us discuss the gauge field action first. We consider the following class
of improved actions: 
\be
\begin{split}
S_G & = \frac{6}{g^2} \,\,\Bigg[c_0 
\sum_{\rm plaquette} \frac{1}{3}\, {\rm Re\, Tr\,}(1-U_{\rm plaquette})
\, + \, c_1 \sum_{\rm rectangle} \frac{1}{3}\,{\rm Re \, Tr\,}(1- U_{\rm
rectangle}) \\
&\phantom{=}  + \, c_2 \,\sum_{\rm chair} \frac{1}{3}\,{\rm Re\, Tr\,}(1- U_{\rm chair})
\, + \, c_3 \sum_{\rm parallelogram} \frac{1}{3}\,{\rm Re \,Tr\,}(1-
U_{\rm parallelogram})\Bigg]\,,
\end{split}
\label{gluonaction}
\ee
where $U_{\rm plaquette}$ is the standard plaquette, while the remaining $U$'s
cover all possible closed loops of link matrices of length six along the edges
of the hypercubes, as indicated in Fig.~{\ref{cube}}.  
 It is customary to impose the normalisation condition
\begin{equation}
c_0+ 8 c_1 + 16 c_2 + 8 c_3=1\,.
\label{crelations}
\end{equation}
In Section~\ref{TIsection} we will also consider actions which fulfill this
condition only in the limit $g \rightarrow 0$, which is sufficient to ensure
the correct continuum limit. Calling the coefficient in front of the plaquette
part the lattice coupling $\beta$, we have the obvious relation 
\be
\beta = \frac{6}{g^2}\, c_0 \,.
\label{defbeta}
\ee
If the improvement is performed only on-shell, one of the six-link
contributions can be set to zero~\cite{Luscher:1985zq}. In general, the
parameters $c_i$ are chosen according to an approximate renormalisation group
(RG) analysis or using tadpole improved perturbation theory.

For perturbation theory expansions, we write the $SU(3)$ link matrices as
\be
U_{x,\mu}=\exp\left[{\rm i}g a A_\mu\Big(x+\frac{a}{2}\hat{\mu}\Big)\right] \,,
\ee
where $A_\mu = T^a A_\mu^a$, $T^a$ being the generators of the Lie algebra,
and perform a weak coupling expansion in $g$. The calculations are carried out
in a general covariant gauge, distinguished by the gauge parameter $\xi =
1-\alpha$. Landau gauge corresponds to $\xi=1$, while the Feynman gauge
corresponds to $\xi=0$. The gluon propagator (in the infinite volume) is
obtained from the lowest order expansion of (\ref{gluonaction}):
\be
S_{\rm G}^{(0)} = \frac{1}{2}\int_{-\pi/a}^{\pi/a} \frac{d^4k}{(2\pi)^4}
\sum_{\mu\nu} 
A_\mu^a(k)\left[G_{\mu\nu}(k)-\frac{\xi}{\xi-1}\hat{k}_\mu\hat{k}_\nu\right]
A_\nu^a(-k)\,,
\label{gluon0}
\ee
where
\be
G_{\mu\nu}(k) = \hat{k}_\mu\hat{k}_\nu + \sum_\rho \left(
\hat{k}_\rho^2 \delta_{\mu\nu} - \hat{k}_\mu\hat{k}_\rho \delta_{\rho\nu}
\right)  \, d_{\mu\rho}
\ee
and
\be
d_{\mu\nu}= \left(1-\delta_{\mu\nu}\right)
\left[C_0 -
C_1 \, a^2 \hat{k}^2 -  C_2 \, a^2( \hat{k}_\mu^2 + \hat{k}_\nu^2)
\right]
\,, \quad
\hat{k}_\mu = \frac{2}{a}\sin\frac{ak_\mu}{2}\,, \quad
        \hat{k}^2 = \sum_\mu \hat{k}_\mu^2  \,.
\label{abbrev}
\ee
The coefficients $C_i$ are related to the coefficients $c_i$ of the
improved action by 
\be
 C_0 = c_0 + 8 c_1 + 16 c_2 + 8 c_3 \,, \,\,\,
C_1 = c_2 + c_3\,, \,\,\, C_2 = c_1 - c_2 - c_3 \,.
\label{C1-C2}
\ee
 (We will usually work with actions in which $C_0 = 1$; see
 eq.~(\ref{crelations}).)  

In lattice momentum space the gluon propagator $D_{\mu\nu}(k)$ is given by the
set of linear equations
\be
\sum_\rho \left[G_{\mu\rho}(k) -
  \frac{\xi}{\xi-1}\hat{k}_\mu\hat{k}_\rho\right]  D_{\rho\nu}(k) =
        \delta_{\mu\nu} \,.
\label{ortho}
\ee
For the plaquette action the gluon propagator reads
\be
 D_{\mu\nu}^{\rm \,plaquette}(k) = \frac{1}{\hat{k}^2}\left( \delta_{\mu\nu} - 
 \xi\, \frac{\hat{k}_\mu \hat{k}_\nu}{\hat{k}^2}\right) \,.
\label{wilsongluon}
\ee
Expressions for the more general action can be found in the
literature~\cite{Iwasaki:1983ck,Weisz:1982zw}. We will use dimensional
regularisation to regularise the loop integrals. Thus we need to know
$D_{\mu\nu}$ for arbitrary dimensions. In general, $D_{\mu\nu}$ can be
given in closed form only for integer dimensions. Only for the special case
$C_2 = 0$ can we derive explicit expressions for arbitrary dimensions. 
A way out is to split the gluon propagator into two parts: a singular part,
which can easily be extended to arbitrary dimensions,
and a finite part, which does not need to be regularised.
This is achieved by writing
\be
D_{\mu\nu} = C_0^{-1}\,D_{\mu\nu}^{\rm \, plaquette} + \Delta D_{\mu\nu} \,.
\label{fullprop}
\ee
It is readily seen that $D_{\mu\nu}$ and
 $C_0^{-1}\,D_{\mu\nu}^{\rm \, plaquette}$ have
the same infrared singularity. The finite part of the gluon propagator $\Delta
D_{\mu\nu}$ is obtained by solving (\ref{ortho}) for the difference
$D_{\mu\nu} - C_0^{-1}\,D_{\mu\nu}^{\rm \, plaquette}$
 (in four dimensions). It can be written 
\be
\Delta D_{\mu\nu } =  \delta_{\mu\nu} 
\sum_{n=0}^4 D_n(k,C)\,\hat{k}_\mu^{2n}
 + \sum_{n=0}^4 \sum_{m=0}^{4-n} 
 D_{m,n}(k,C)\,\hat{k}_\mu^{2m+1}\,\hat{k}_\nu^{2n+1}\,.
\label{DeltaD}
\ee
The scalars $D_n$ and $D_{m,n}=D_{n,m}$ are rational functions of
$\hat{k}_\mu$ and $C_i$. They are listed in Appendix A for
 the case $C_0=1$.

The final results cannot be expressed in analytic form (as a function of
$C_i$) anymore. We therefore have to make a choice concerning the
parameters. We restrict ourselves to the most popular actions in this class:

\noindent
\underline{Tree-level Symanzik}~\cite{Symanzik:1983dc}

\be
c_1 = -1/12,\; c_2  = c_3 = 0\,.
\ee

\noindent
\underline{Tadpole improved L\"uscher-Weisz
  (TILW)}~\cite{Luscher:1984xn,Luscher:1985zq,Alford:1995hw}  

\be
c_2 = 0\,.
\ee
Once we have chosen $\beta$, the other parameters are
fixed~\cite{Alford:1995hw}: 
\be
\frac{c_1}{c_0}=-\frac{(1+0.4805\,\alpha)}{20\,u_0^2}\,,\quad 
\frac{c_3}{c_0}=-\frac{0.03325\,\alpha}{u_0^2}\,, \quad
\frac{1}{c_0}=1+8\left(\frac{c_1}{c_0}+\frac{c_3}{c_0}\right)\,, 
\ee
where
\be
u_0=\left(\frac{1}{3}\, {\rm Tr}\,\langle U_{\rm
    plaquette}\rangle\right)^{\frac{1}{4}}
\label{u0}
\ee
and 
\be
\alpha = -\frac{\log(u_0^4)}{3.06839}\,.   
\vspace*{0.25cm}
\ee
We have picked 
the following values~\cite{Gattringer:2001jf}:
\vspace*{0.5cm}
\begin{equation*}
\begin{tabular}{c|c|c}
  $\beta$  & $c_1$ & $c_3$  \\
  \hline
  8.60  &$-0.151791$&$-0.0128098$  \\ [0.7ex]
  8.45  &$-0.154846$&$-0.0134070$  \\ [0.7ex]
  8.30  &$-0.159128$&$-0.0142442$  \\ [0.7ex]
  8.20  &$-0.161827$&$-0.0147710$  \\ [0.7ex]
  8.10  &$-0.165353$&$-0.0154645$  \\ [0.7ex]
  8.00  &$-0.169805$&$-0.0163414$  
  \end{tabular}
\vspace*{0.5cm}
\label{tabc}
\end{equation*}
which corresponds to lattice spacings $a = 0.084 - 0.136$ fm. These $\beta$
values are currently being employed in quenched Monte Carlo simulations,
or most probably will be in the near future. 

\noindent
\underline{Iwasaki}~\cite{Iwasaki:1983ck}

\be
c_1 = -0.331,\; c_2 = c_3=0\,.
\ee

\noindent
\underline{DBW2}~\cite{Takaishi:1996xj}

\be
c_1 = -1.4086,\; c_2 = c_3=0\,.
\ee
 In all these cases $c_0$ is chosen to satisfy eq.~(\ref{crelations}). 

Which action is the best one? They all reduce or eliminate $O(a^2)$ corrections
-- at the tree level, the one-loop level, or beyond. In Section 4 we compare
the actions with regard to their perturbative merits. 

The action for massless overlap fermions is given by
\be
S_F=\bar{\psi} D_N \psi\,.
\ee
The Neuberger-Dirac operator $D_N$ was already given in eq.~(\ref{over}). The 
Wilson-Dirac operator $D_W$ reads
\be
D_W = \frac{1}{2} \Big[ \gamma_\mu (\nabla^\star_\mu + \nabla_\mu)
- a r \nabla^\star_\mu \nabla_\mu \Big] \,,
\label{DW}
\ee
where $\nabla_\mu$ is the lattice forward covariant derivative:
\be
\nabla_\mu\, \psi (x) =
\frac{1}{a} \Big[ U_{x,\mu}\, \psi(x+a\hat{\mu}) - \psi (x) \Big] \,.
\ee
For $0 <\rho < 2r$ the correct spectrum of massless fermions without
doubling is obtained~\cite{n}.

The lattice Feynman rules for overlap fermions, being originally
derived in~\cite{ky,ikny}, are collected in Appendix A. The regularisation of
the infrared divergences  
follows~\cite{Gockeler:1996hg}, which was adapted from~\cite{Kawai:1980ja}.
The calculations are done analytically as far as possible. To do so,
we have 
extended our {\sl Mathematica} programme package, which we developed
originally for Wilson~\cite{Gockeler:1996hg} and clover
fermions~\cite{Capitani:2000xi}, to overlap fermions with improved gauge
actions. 

We performed several tests to check the code. Operators with gamma structures
$\Gamma$ and $\Gamma \gamma_5$ must give the same result. We have performed
calculations independently for both sets of gamma matrices. Although the
analytic expressions looked very much different, we found complete
agreement. Another condition is that all terms $\propto 1/a$ cancel.
Furthermore, it is expected~\cite{Capitani:2000wi} that the gauge
dependent parts of the one-loop lattice integrals are independent of the
particular choice of fermion propagator (be it Wilson or overlap), as a
consequence of the Ward identities. This has been verified numerically and
analytically. In Appendix B we give an analytic proof.

\section{Renormalisation}
\label{sec:3new}

To obtain finite answers, the lattice operators $\mathcal{O}(a)$ must, in
general, be renormalised. Ignoring operator mixing, we define renormalised
operators by
\begin{equation}
\mathcal{O}^{\mathcal{S}}(\mu) = Z_{\mathcal{O}}^{\mathcal{S}}(a,\mu)
\mathcal{O}(a)\,,    
\end{equation}
where $\mathcal{S}$ denotes the renormalisation scheme. The renormalisation
constants $Z_{\mathcal{O}}$ are often defined in the $MOM$
scheme first by computing the gauge fixed quark propagator $S_N$ and the
amputated Green function $\Lambda_{\mathcal{O}}$ of the operator
$\mathcal{O}$:  
\begin{eqnarray}
Z_\psi^{MOM}(a,\mu)\; S_N\; \big|_{p^2=\mu^2}
&=&  S^{\rm tree} \,, \\[0.3em]
\frac{Z^{MOM}_{\cal O}(a,\mu) }{Z^{MOM}_\psi(a,\mu)}\;
\Lambda_{\cal O}\; \big|_{p^2=\mu^2}
&=& \Lambda_{\cal O}^{\rm tree}
+{\rm \ other\  Dirac\  structures}\,.
\label{ZOlat}
\end{eqnarray}
(Note that $Z_\psi = 1/Z_2$.) The renormalisation constants can be converted
to the $\MS$ scheme,
\begin{equation}
Z_{\psi,\mathcal{O}}^{\MS}(a,\mu) =  Z_{\psi,\mathcal{O}}^{\MS,MOM} 
Z_{\psi,\mathcal{O}}^{MOM}(a,\mu)\,,
\end{equation}
where $Z_{\psi,\mathcal{O}}^{\MS,MOM}$ is calculable in continuum perturbation
theory.

We shall restrict all our numerical calculations to the case $r=1$. The
optimal choice for $\rho$ appears to be $\rho \approx 1.4$. We will present
results for $\rho= 1.3$, 1.35, 1.4, 1.45 and 1.5, which should cover the most
interesting region. Any other value in this region may be obtained by inter- or
extrapolation.

\noindent
{\it Self-energy and wave function renormalisation}

\begin{figure}[!tb]
 \begin{center}
  \begin{picture}(400,100)(0,0)
    \SetOffset(30,-10)
    \GlueArc(100,20)(30,0,180){5}{10}
    \ArrowLine(10,20)(70,20)
    \ArrowLine(70,20)(130,20)
    \ArrowLine(130,20)(190,20)
    \Text(30,5)[bl]{$p$}
    \Text(90,5)[bl]{$p+k$}
    \Text(160,5)[bl]{$p$}
    \Text(100,75)[tl]{$k$}
    \ArrowLine(210,20)(260,20)
    \ArrowLine(260,20)(320,20)
    \GlueArc(260,45)(20,-90,270){5}{13}
    \Text(235,5)[bl]{$p$}
    \Text(285,5)[bl]{$p$}
    \Text(260,80)[bl]{$k$}   
  \end{picture}
 \end{center}
\caption{One-loop diagrams contributing to the quark self-energy.}
 \label{fig1}
\end{figure}
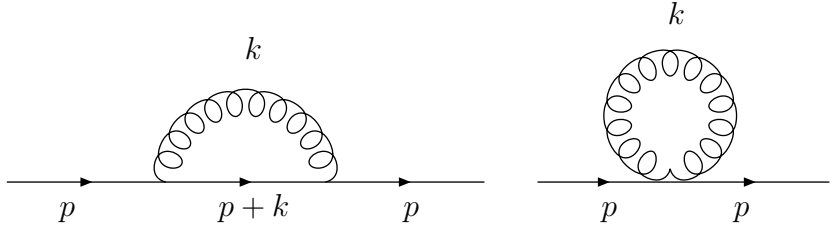

Let us consider the massless quark propagator $S_N$ first. The inverse of
$S_N$ can be written
\begin{equation}
  S_N^{-1} = {\rm i}\pslash \left( 1 - \; 
  \frac{g^2\,C_F}{16 \pi^2}\, \Sigma_1 \right)\,,
  \label{qprop}
\end{equation}
with $C_F=4/3$. The diagrams that contribute to $\Sigma_1$ to one-loop order
are shown in Fig.~\ref{fig1}. The integral to be evaluated
is~\cite{Gockeler:1996hg}
\begin{equation}
\frac{\Sigma_1}{16\pi^2 } = \int\frac{d^{\rm D}k}{(2\pi)^{\rm D}}
\sum_{\mu\nu}\Bigl[V_{1\,\mu}(p,p+k)\,S_N(p+k)\, 
  V_{1\,\nu}(p+k,p) + V_{2\,\mu\nu}(p,p,k,k)\Bigr]\,D_{\mu\nu}(k) \,.
\end{equation}
Explicit expressions for the propagator and vertex functions are given in
Appendix A. Putting everything together, we finally obtain 
\begin{equation}
  \Sigma_1(a,p)=  (1-\xi) \, \log(a^2 p^2) + 4.79201 \, \xi + b_\Sigma\,.
\label{Sigma}
\end{equation}
The results for $b_\Sigma$ are given in Table~\ref{tab1} for the gauge field
actions and $\rho$ parameters mentioned. The result for the plaquette action
($c_0=1$, $c_1 = c_2 = c_3 = 0$) agrees with
Refs.~\cite{Alexandrou:2000kj,Capitani:2000wi}. The gauge dependent part of
(\ref{Sigma}) is the same for all actions:
\begin{equation}
(F_0 -\gamma_E+1) \, \xi = 4.79201 \, \xi \,,
\label{gaugedep}
\end{equation}
with
\begin{equation}
F_0=4.3692252338748\,, \quad \gamma_E=0.57721566\,.
\end{equation}
We explain the reason for this observation in Appendix B.
For the wave function renormalisation constant we then obtain
\begin{equation}
\begin{split}
 Z_\psi^{MOM}(a,\mu) &=  1 - \frac{g^2 C_F}{16\pi^2}\,
 \Sigma_1(a,\mu)  \\
& = 1 -\frac{g^2 C_F}{16\pi^2} \,
\left[ 2 (1-\xi) \log (a \mu)+ 4.79201 \, \xi + b_\Sigma \right] \,.
\label{zpsimom}
\end{split}
\end{equation}
In the $\MS$ scheme this becomes
\begin{equation}
 Z_\psi^{\MS}(a,\mu)  = 1 -\frac{g^2 C_F}{16\pi^2} \,
\left[ 2 (1-\xi) \log (a \mu)+ 3.79201 \, \xi + b_\Sigma +1\right] \,.
\label{zpsi}
\end{equation}
For Wilson fermions (with $r=1$), on the other hand, we have
\begin{equation}
   Z_{\psi,{\rm Wilson}}^{\MS}(a,\mu) = 1 -\frac{ g^2 C_F}{16 \pi^2}  \,
   \left[2 (1-\xi) \log (a \mu)+ 3.79201 \, \xi
   +12.8524 \right] \,.
\label{zpsiw}
\end{equation}
We observe that the constant, gauge independent terms in eqs.~(\ref{zpsi}) and
(\ref{zpsiw}) have opposite sign. So for $\mu = 1/a$ the overlap $Z_\psi$ is
greater than one, while $Z_{\psi,{\rm Wilson}}$ is less than one.

\begin{table}[!t]
  \begin{center}
\begin{tabular}{lrrrrr}
\hline
Action  & \multicolumn{4}{l}{\rule[-3mm]{0mm}{8mm}$\rho$} \\ \cline{2-6}\\[-1.25ex]
 % \\ %\cline{2-6}
 & $1.30\phantom{0}$ & $1.35\phantom{0}$ & $1.40\phantom{0}$ &
 $1.45\phantom{0}$ & $1.50\phantom{0}$\\
\hline & & & & & \\[-1.25ex]
 Plaquette           & $-26.400$ & $-25.031$ & $-23.766$ & $-22.592$ & $-21.501$ \\[0.7ex]
 Symanzik            & $-21.774$ & $-20.713$ & $-19.732$ & $-18.823$ & $-17.978$\\ [0.7ex]
 TILW, $\beta$=8.60 & $-19.216$ & $-18.329$ & $-17.509$ & $-16.749$ & $-16.044$\\ [0.7ex]
 TILW, $\beta$=8.45 & $-19.124$ & $-18.243$ & $-17.429$ & $-16.675$ & $-15.974$\\ [0.7ex]
 TILW, $\beta$=8.30 & $-18.997$ & $-18.124$ & $-17.319$ & $-16.572$ & $-15.878$\\ [0.7ex]
 TILW, $\beta$=8.20 & $-18.918$ & $-18.051$ & $-17.25$ & $-16.508$ & $-15.819$\\ [0.7ex]
 TILW, $\beta$=8.10 & $-18.817$ & $-17.957$ & $-17.163$ & $-16.426$ & $-15.743$\\ [0.7ex]
 TILW, $\beta$=8.00 & $-18.692$ & $-17.841$ & $-17.054$ & $-16.325$ & $-15.648$\\ [0.7ex]
 Iwasaki             & $-15.960$ & $-15.295$ & $-14.681$ & $-14.112$ & $-13.584$\\ [0.7ex]
 DBW2                & $-10.373$ & $-10.096$ & $-9.841$ & $-9.605$ & $-9.386$\\ [0.5ex] \hline
\end{tabular}
  \end{center}
  \caption{The contribution $b_\Sigma$ to the self-energy for the various
    actions and parameters.}
  \label{tab1}
\end{table}

\newpage
\noindent
{\it Quark bilinears}

\begin{figure}[!t]
 \begin{center}
  \begin{picture}(400,130)(0,0)
   \SetOffset(130,-10)
   \ArrowLine(10,10)(36,50)
   \ArrowLine(36,50)(70,100)
   \ArrowLine(70,100)(104,50)
   \ArrowLine(104,50)(130,10)
   \GCirc(70,100){7}{0}
   \Gluon(36,50)(104,50){5}{8}
   \Text(10,30)[bl]{$p$}
   \Text(20,70)[bl]{$p+k$}
   \Text(100,70)[bl]{$p+k$}
   \Text(125,30)[bl]{$p$}
   \Text(70,30)[bl]{$k$}
   \Text(70,115)[bl]{$\Gamma^X$}
  \end{picture}
 \end{center}
\caption{One-loop vertex diagram contributing to the amputated Green
   functions.} 
 \label{fig3}
\end{figure}
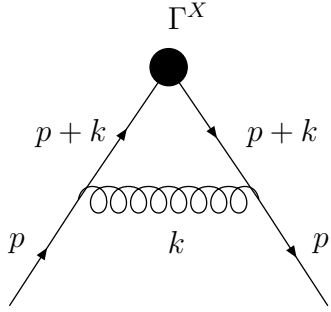

Let us consider local operators of the form
\begin{equation}
  {\cal O}_X = \bar{\psi} (x) \Gamma^X \psi(x)
\end{equation}
with 
\begin{equation}
\Gamma^S=\bbbone\,, \  \Gamma^P=\gamma_5\,, \ \Gamma^V=\gamma_\mu\,,  \
\Gamma^A=\gamma_\mu\gamma_5 \,, \
\Gamma^T=\sigma_{\mu\nu} \gamma_5 \,,
\label{GammaDef}
\end{equation}
i.e. $X=S$, $P$, $V$, $A$ and $T$. As the operators are local, operator
tadpole and cockscomb diagrams do not contribute~\cite{Gockeler:1996hg}. 
This leaves us to compute the vertex diagram shown in Fig.~\ref{fig3}.
We denote the amputated Green function of the operator $\mathcal{O}_X$ by
$\Lambda^X$. Thus we have to evaluate the integral
\begin{equation}
\begin{split}
\Lambda^X = \Gamma^X + g^2 C_F \int \frac{d^{\rm D}k}{(2\pi)^{\rm D}}
\sum_{\mu\nu} \,&V_{1\,\mu}(p,p+k)\, S_N(p+k) \\
 &\times \,{\Gamma}^X\, S_N(p+k)\, V_{1\,\nu}(p+k,k) \,
 D_{\mu\nu}(k) \,. 
\end{split}
\end{equation}
The final result is
\begin{eqnarray}
 \Lambda^{S,P} &=& \{\bbbone,\gamma_5\} + \frac{g^2 C_F}{16\pi^2}
 \left[(\xi-4) \, \log(a^2 p^2) - 5.79201 \xi + b_{S,P} \right]\,
 \{\bbbone,\gamma_5\} \,, \\ [0.3em]
 \Lambda_\mu^{V,A} &=& \{\gamma_\mu,\gamma_\mu\gamma_5\} + \frac{g^2
 C_F}{16\pi^2} \biggl\{ \gamma_\mu \left[ (\xi-1) \, \log(a^2 p^2) - 4.79201
 \xi +   b_{V,A}  \right] \nonumber \\ [0.3em]
 & & \hspace*{6.425cm} -\, 2(1-\xi)\,   \frac{p_\mu \pslash} {p^2}
 \biggr\}\,  \{\bbbone,\gamma_5\} \,, \\ %[0.3em]
 \Lambda_{\mu\nu}^T &=& \sigma_{\mu\nu} \gamma_5  + \frac{g^2 C_F}{16\pi^2}
 \left[ \xi \log(a^2 p^2) - 3.79201  \xi +  b_{T} \right] \,
 \sigma_{\mu\nu} \gamma_5 \,.
\label{3PF}
\end{eqnarray}

\clearpage
\begin{table}[!t]
\begin{center}
\begin{tabular}{lrrrrr}
\hline
Action  & \multicolumn{4}{l}{\rule[-3mm]{0mm}{8mm}$\rho$} \\ \cline{2-6}
 \\[-1.25ex] %\cline{2-6}
 & $1.30\phantom{0}$ & $1.35\phantom{0}$ & $1.40\phantom{0}$ &
 $1.45\phantom{0}$ & $1.50\phantom{0}$\\
\hline & & & & & \\[-1.25ex]
 Plaquette           & $11.083$ & $11.214$ & $11.343$ & $11.472$ & $11.600$ \\[0.7ex]
 Symanzik            & $10.621$ & $10.739$ & $10.856$ & $10.972$ & $11.086$\\ [0.7ex]
 TILW, $\beta$=8.60 & $10.275$ & $10.385$ & $10.493$ & $10.599$ & $10.705$\\ [0.7ex]
 TILW, $\beta$=8.45 & $10.261$ & $10.370$ & $10.478$ & $10.584$ & $10.689$\\ [0.7ex]
 TILW, $\beta$=8.30 & $10.241$ & $10.350$ & $10.457$ & $10.563$ & $10.667$\\ [0.7ex]
 TILW, $\beta$=8.20 & $10.229$ & $10.337$ & $10.444$ & $10.549$ & $10.653$\\ [0.7ex]
 TILW, $\beta$=8.10 & $10.213$ & $10.321$ & $10.427$ & $10.532$ & $10.636$\\ [0.7ex]
 TILW, $\beta$=8.00 & $10.192$ & $10.300$ & $10.406$ & $10.510$ & $10.613$\\ [0.7ex]
 Iwasaki             & $9.641$ & $9.736$ & $9.830$ & $9.922$ & $10.012$\\ [0.7ex]
 DBW2                & $7.457$ & $7.515$ & $7.571$ & $7.626$ & $7.679$\\ [0.5ex] \hline
\end{tabular}
  \end{center}
  \caption{The contribution $b_{S,P}$ to $\Lambda^{S,P}$ for various actions
    and parameters.} 
  \label{tabbsp}
\vspace*{0.5cm}
\begin{center}
 \begin{tabular}{lrrrrr}
\hline
Action  & \multicolumn{4}{l}{\rule[-3mm]{0mm}{8mm}$\rho$} \\ \cline{2-6}
 \\[-1.25ex] %\cline{2-6}
 & $1.30\phantom{0}$ & $1.35\phantom{0}$ & $1.40\phantom{0}$ &
 $1.45\phantom{0}$ & $1.50\phantom{0}$\\
\hline & & & & & \\[-1.25ex]
 Plaquette           & $6.343$ & $6.345$ & $6.348$ & $6.351$ & $6.354$ \\[0.7ex]
 Symanzik            & $6.332$ & $6.333$ & $6.335$ & $6.338$ & $6.340$\\ [0.7ex]
 TILW, $\beta$=8.60 & $6.325$ & $6.327$ & $6.329$ & $6.330$ & $6.332$\\ [0.7ex]
 TILW, $\beta$=8.45 & $6.325$ & $6.327$ & $6.328$ & $6.330$ & $6.332$\\ [0.7ex]
 TILW, $\beta$=8.30 & $6.325$ & $6.326$ & $6.328$ & $6.330$ & $6.332$\\ [0.7ex]
 TILW, $\beta$=8.20 & $6.325$ & $6.326$ & $6.328$ & $6.330$ & $6.331$\\ [0.7ex]
 TILW, $\beta$=8.10 & $6.324$ & $6.326$ & $6.328$ & $6.329$ & $6.331$\\ [0.7ex]
 TILW, $\beta$=8.00 & $6.324$ & $6.326$ & $6.327$ & $6.329$ & $6.331$\\ [0.7ex]
 Iwasaki             & $6.316$ & $6.318$ & $6.319$ & $6.320$ & $6.321$\\ [0.7ex]
 DBW2                & $6.302$ & $6.302$ & $6.303$ & $6.303$ & $6.304$\\ [0.5ex] \hline
\end{tabular}
  \end{center}
  \caption{The contribution $b_{V,A}$ to $\Lambda^{V,A}$ for various actions
    and parameters.} 
  \label{tabbva}
\end{table}

\clearpage
\begin{table}[!t]
\begin{center}
 \begin{tabular}{lrrrrr}
\hline
Action  & \multicolumn{4}{l}{\rule[-3mm]{0mm}{8mm}$\rho$} \\ \cline{2-6}
 \\[-1.25ex] %\cline{2-6}
 & $1.30\phantom{0}$ & $1.35\phantom{0}$ & $1.40\phantom{0}$ &
 $1.45\phantom{0}$ & $1.50\phantom{0}$\\
\hline & & & & & \\[-1.25ex]
 Plaquette           & $4.096$ & $4.056$ & $4.016$ & $3.977$ & $3.938$ \\[0.7ex]
 Symanzik            & $3.958$ & $3.914$ & $3.871$ & $3.828$ & $3.785$\\ [0.7ex]
 TILW, $\beta$=8.60 & $3.851$ & $3.804$ & $3.759$ & $3.713$ & $3.668$\\ [0.7ex]
 TILW, $\beta$=8.45 & $3.846$ & $3.800$ & $3.754$ & $3.708$ & $3.664$\\ [0.7ex]
 TILW, $\beta$=8.30 & $3.840$ & $3.793$ & $3.747$ & $3.702$ & $3.657$\\ [0.7ex]
 TILW, $\beta$=8.20 & $3.836$ & $3.789$ & $3.743$ & $3.698$ & $3.652$\\ [0.7ex]
 TILW, $\beta$=8.10 & $3.831$ & $3.784$ & $3.738$ & $3.692$ & $3.647$\\ [0.7ex]
 TILW, $\beta$=8.00 & $3.825$ & $3.778$ & $3.731$ & $3.686$ & $3.640$\\ [0.7ex]
 Iwasaki             & $3.651$ & $3.601$ & $3.551$ & $3.501$ & $3.452$\\ [0.7ex]
 DBW2                & $2.943$ & $2.881$ & $2.819$ & $2.759$ & $2.698$\\ [0.5ex] \hline
\end{tabular}
  \end{center}
  \caption{The contribution $b_T$ to $\Lambda^T$ for various actions and
    parameters.} 
  \label{tabbt}
\end{table}

The finite terms $b_{S,P}$, $b_{V,A}$ and $b_{T}$ are collected in
Tables~\ref{tabbsp},~\ref{tabbva} and~ \ref{tabbt}. For the renormalisation
constants we then obtain, using (\ref{ZOlat}) and (\ref{zpsimom}),
\begin{eqnarray}
  Z_{S,P}^{MOM}(a,\mu) & = &
  1 - \frac{g^2 C_F}{16 \pi^2} \left[-6 \log(a\mu)- \xi + b_{S,P} + b_\Sigma
  \right] \,, \\
  Z_{V,A}^{MOM}(a,\mu) & = &
  1 - \frac{g^2 C_F}{16 \pi^2} \left(b_{V,A} + b_\Sigma
  \right) \,, \\
  Z_{T}^{MOM}(a,\mu) & = &
  1 - \frac{g^2 C_F}{16 \pi^2}  \left[2\log(a\mu)+ \xi+ b_T
  +b_\Sigma  \right] \,. 
\end{eqnarray}
In the $\MS$ scheme the renormalisation constants read
\begin{eqnarray}
  Z_{S,P}^{\MS}(a,\mu) & = &  1 - \frac{g^2 C_F}{16 \pi^2} \left[-6
  \log(a\mu)- 5 + b_{S,P} + b_\Sigma  \right] \,, \\
  Z_{V,A}^{\MS}(a,\mu) & = &  1 - \frac{g^2 C_F}{16 \pi^2} \left(b_{V,A} +
  b_\Sigma  \right)\,, \\
  Z_T^{\MS}(a,\mu) & = &  1 - \frac{g^2 C_F}{16 \pi^2}
  \left[2\log(a\mu)+ 1 + b_T +b_\Sigma  \right] \,.
\end{eqnarray}
The conversion factors $Z_{\mathcal{O}}^{\MS,MOM}$ are universal, and are the
same as for the plaquette action and Wilson fermions~\cite{Gockeler:1996hg}.

\section{Tadpole improvement \label{TIsection}}

The appearance of gluon tadpoles in lattice perturbation theory make the bare
coupling constant $g$ into a poor expansion parameter. It was
proposed~\cite{Lepage:1992xa} that the perturbative series should be
rearranged in order to get rid of these contributions. Tadpole improvement is
a technique for summing, to all orders, the numerically large perturbative
contributions arising from tadpole diagrams. It is implemented by a mean field
renormalisation of the link matrices, $U_{x,\mu} \rightarrow U_{x,\mu}/u_0$,
where $u_0$ 
is the mean value of the link, defined to be the fourth root of the
expectation value of the plaquette `measured' in Monte Carlo simulations,
as given in eq.~(\ref{u0}). In this Section we apply tadpole improvement,
better called mean field improvement, to our results, and to operators
involving overlap fermions in general. 

    Our situation is more complicated than the standard case
 for two reasons: firstly we are using overlap fermions rather than
 Wilson or clover fermions, and secondly we are using gauge actions 
 which are more complicated than the basic plaquette action. 
 Both these changes require some thought.

    First, let us run briefly through the procedure for the case of
 Wilson fermions. Tadpole improved renormalisation constants are defined by
\begin{equation}
Z_\mathcal{O}^{TI} = Z_\mathcal{O}^{MF}
\left(\frac{Z_\mathcal{O}}{Z_\mathcal{O}^{MF}}\right)_{\rm pert}\,,
\label{zti}
\end{equation}
where $Z_\mathcal{O}^{MF}$ is the mean field approximation of
$Z_\mathcal{O}$, while the second factor on the r.h.s. is computed in
perturbation theory. We find $Z_{\mathcal{O},{\rm Wilson}}^{MF}$ by
 looking at the operator Green function and quark propagator in the 
 mean field approximation. The mean field result for the amputated
 Green function for an operator with $n_D$ covariant derivatives is 
 \begin{equation}
\Lambda_\mathcal{O}^{MF} = u_0^{n_D} \Lambda_\mathcal{O}^{\rm tree}\,.
 \label{mfLambda}
\end{equation}
 This does not depend on the choice of fermion action. 
 In the mean field approximation the inverse quark propagator 
 for massless Wilson fermions is 
 \begin{equation}
 (S^{MF}_{Wilson})^{-1} = {\rm i} \pslash \, u_0 + O(a)\,,
 \label{Swils}
 \end{equation}
 giving 
  \begin{equation}
 Z_{\psi,Wilson}^{MF} = u_0 \,. 
 \label{zpsimfw} 
 \end{equation}
 Combining eqs.~(\ref{mfLambda}) and (\ref{zpsimfw}) gives
 the familiar result
\begin{equation}
Z_{\mathcal{O},{\rm Wilson}}^{MF} = u_0^{1-n_D} \,.
\label{zmfw}
\end{equation}

   The result changes when we consider overlap fermions. 
 The mean field result for the amputated Green functions,
 eq.~(\ref{mfLambda}), is  unchanged. However,  $Z_\psi$ for
 overlap fermions is no longer given by~(\ref{zpsimfw}), 
which will lead to changes in $Z_\mathcal{O}$ as well. To find 
  $Z_\psi$ we need to calculate the overlap operator in the mean field
 approximation. The mean field result for the Wilson-Dirac operator
 in momentum space is 
\begin{equation}
\begin{split}
 D_W^{MF} &= \frac{1}{a} \left(
  u_0  \sum_\mu {\rm i}\, \gamma_\mu \sin a p_\mu + 4 r
 - u_0\,  r \sum_\mu \cos a p_\mu \right) \\
 &= {\rm i} \pslash \; u_0 + \frac{4 r }{a} (1 -u_0) + \frac{r}{2} a p^2 u_0
 +O(a^2) \,.
\end{split}
\end{equation}
Inserting this result into the Neuberger-Dirac operator (\ref{over}) gives
\begin{equation}
 D_N^{MF} = \frac{\rho \; u_0}{\rho -4 r (1- u_0)} \left[{\rm i} \pslash 
 + \frac{1}{2}\, a\, p^2 \, \frac{u_0}{\rho -4r(1-u_0)} + O(a^2)\right]\,.
\label{finverse}
\end{equation}
We can now compute $Z_\psi^{MF}$:
\begin{equation}
 Z_\psi^{MF} =  \frac{\rho \; u_0}{\rho -4 r (1- u_0)} \,.
 %= 1 - \ggcf \left(1 - \frac{4 r}{\rho}\right) k_u + O(g^4).
 \label{Zpsimf} 
\end{equation}
If we combine this result with eq.~(\ref{mfLambda}), we finally obtain
\begin{equation}
 Z_{\cal O}^{MF} =  \frac{\rho \; u_0^{1-n_D} }{\rho -4 r (1- u_0)} \,.
 %= 1 - \ggcf \left(1 -n_D - \frac{4 r}{\rho}\right) k_u + O(g^4).
 \label{ZOmf} 
\end{equation}
It is required that $\rho > 4r(1-u_0)$. 
 The reason for this inequality is easy to understand.
 Reverting temporarily to writing the fermion matrix
 in terms of a hopping parameter $\kappa$,
 the $\rho$ definition in eq.~(\ref{over}) is equivalent to
 $\rho = 4 r  -1/(2 \kappa)$. To really have a negative mass in $X$, so
 that the overlap procedure works, requires $\kappa > \kappa_c$,
 which implies $\rho > 4 r - 1/(2 \kappa_c) $. In the mean
 field approximation $\kappa_c = 1/(8 r u_0)$, leading to the
 inequality $\rho > 4r(1-u_0)$. 
 In other words, the additive renormaliation of
 mass, which occurs with Wilson fermions, means that in the interacting case
 $\rho > 0$ is not enough to cause $X$ to have a negative mass. We need
 $\rho$ to be large enough to overcome this additional mass.

  Note that $Z_\psi^{MF}$ will be larger than 1 for reasonable values
 of $\rho$, while in the Wilson case $Z_{\psi,Wilson}^{MF} < 1$. 
 Our mean field calculation gives an explanation of the observation
 in Section 3, that the one-loop $b_\Sigma$'s have the opposite
 sign for overlap fermions than for Wilson fermions. 

\begin{table}[!t]
\begin{center}
  \begin{tabular}{|c|c|c|c|c|}
  \hline \rule[-3mm]{0mm}{8mm}
 Action & $\beta$ & $k_u$ & $k_u^{TI}$ & $u_0^4$ \\ \hline & & & & \\[-1.25ex]
 Plaquette &  \phantom{1}6.0\phantom{00} & $\pi^2$ & $\pi^2$  & 0.59368 \\
 [0.7ex]
  Symanzik &   & 0.732524\,$\pi^2$ & & \\  [0.7ex]
 TILW     &  \phantom{1}8.60\phantom{0} & 0.590078\,$\pi^2$ & 0.549643\,$\pi^2$
 &  0.66018 \\ [0.7ex] 
 TILW     &  \phantom{1}8.45\phantom{0} & 0.585039\,$\pi^2$ &
 0.543338\,$\pi^2$ &  0.65176 \\ [0.7ex] 
 TILW     &  \phantom{1}8.30\phantom{0} & 0.578127\,$\pi^2$ &
 0.534971\,$\pi^2$ &  0.64252 \\ [0.7ex] 
 TILW     &  \phantom{1}8.20\phantom{0} & 0.573860\,$\pi^2$ &
 0.529705\,$\pi^2$ &  0.63599 \\ [0.7ex] 
 TILW     &  \phantom{1}8.10\phantom{0} & 0.568378\,$\pi^2$ &
 0.523106\,$\pi^2$ &  0.62894 \\ [0.7ex] 
 TILW     &  \phantom{1}8.00\phantom{0} & 0.561610\,$\pi^2$ &
 0.515069\,$\pi^2$ &  0.62107 \\ [0.7ex]
 Iwasaki  &  \phantom{1}9.485 & 0.420531\,$\pi^2$ & 0.379397\,$\pi^2$ &
 0.67066 \\ [0.7ex]
 Iwasaki  &  \phantom{1}8.026 & 0.420531\,$\pi^2$ & 0.370379\,$\pi^2$ &
 0.59561 \\ [0.7ex]
 DBW2     &  12.745 & 0.153264\,$\pi^2$ & 0.131857\,$\pi^2$ & 0.72759 \\
 [0.7ex]
 DBW2     &  10.662 & 0.153264\,$\pi^2$ & 0.128275\,$\pi^2$ & 0.67332 \\
  [0.7ex] 
  \hline
  \end{tabular}
  \end{center}
  \caption{The parameters $k_u$,  $k_u^{TI}$ and the average plaquette $u_0^4$
 for various 
 actions and $\beta$ values. See eq.~(\ref{defbeta}) for the definition of
 $\beta$. The plaquette values for the TILW, Iwasaki and DBW2 actions are taken
 from~\cite{Gattringer:2001jf}, \cite{Aoki:2002iq} and \cite{Blum},
 respectively.}  
  \label{tabku}
\end{table}

Let us now turn to the calculation of the second, perturbative factor on the
r.h.s. of eq.~(\ref{zti}). Having removed the large tadpole contributions,
we need to re-express the perturbative series in terms of the tadpole improved
parameters, which must satisfy
 \begin{eqnarray}
 \frac{ c_0^{TI} }{ g_{TI}^2 } &=& u_0^4 \; \frac{c_0 }{ g^2 }\,, \\
 \frac{ c_i^{TI} }{ g_{TI}^2 } &=& u_0^6 \; \frac{c_i }{ g^2 } 
  \,, \quad i=1, 2, 3\,,
 \end{eqnarray}
 because the plaquette term in the action is a four-link term, 
 while the other three terms in the action are six-link objects.
 These conditions are not enough to fully fix the tadpole improved
 parameters, because we have four equations with five unknowns, 
 leaving us with some freedom of choice. The simplest choice is to define
 \begin{eqnarray}
 g_{TI}^2 &=&\frac{g^2}{u_0^4} \,, \label{gti}\\
 c_0^{TI} &=& c_0 \label{c0ti} \\
 c_i^{TI} &=& u_0^2 \,c_i \,, \quad i=1, 2, 3\,. \label{cti}
 \end{eqnarray}
 However, it is important to notice that, although this definition makes
 the formulae for the tadpole improved parameters very simple,
 with this choice
 \begin{equation}
  C_0^{TI} \equiv c_0^{TI} + 8  c_1^{TI} + 16  c_2^{TI} + 8  c_3^{TI} 
  \neq 1 \,, 
 \end{equation} 
 which modifies many of the formulae we wrote down in earlier sections.
This means, we have to replace every $g^2$ by $g_{TI}^2$ and every $c_1$, $c_2$
and $c_3$ by $c_1^{TI}$, $c_2^{TI}$ and $c_3^{TI}$, respectively,
 while keeping $c_0$ unchanged. The effect
of introducing tadpole improved coefficients (\ref{cti}) is that the rescaled
gluon propagator remains of the same form as we change $u_0$, thus ensuring
fast convergence. 

For the
perturbative calculation of $Z_{\cal O}^{MF}$ we need to know the perturbative
expansion of   
$u_0$~\cite{Lepage:1992xa,Alford:1995hw}: 
\begin{equation}
u_0 = 1 - \frac{g_{TI}^2 C_F}{16 \pi^2}\, k_u^{TI} \,,
\label{exu}
\end{equation}
where $k_u^{TI}$ is given by the integral
\begin{equation}
k_u^{TI} = 4 \pi^2 a^4 \int_{-\pi/a}^{\pi/a} \frac{d^4 k}{(2 \pi)^4 }
\left[ \hat k_4^2\, D_{11}(k,C^{TI}) -\hat k_1 \hat k_4\,
  D_{14}(k,C^{TI})\right] \,. 
\label{ku}
\end{equation}
Here $D_{11}(k,C^{TI})$ and $D_{14}(k,C^{TI})$ are
components of the gluon 
propagator (\ref{fullprop}), with $C_0=1,\, C_1=c_2+c_3$ and
 $C_2=c_1-c_2-c_3$ being
replaced by  $C_0^{TI} = c_0 + 8  c_1^{TI} + 16  c_2^{TI} + 8  c_3^{TI}$, 
 $C_1^{TI}=u_0^2\, C_1$ and $C_2^{TI}=u_0^2\, C_2$,
respectively. The numerical values of 
$k_u^{TI}$ are given in Table~\ref{tabku} for the various actions. For
comparison, we also give the non-tadpole improved result (i.e. with
coefficients $C_1$ and $C_2$), which we call $k_u$.
 The `measured' values of $u_0^4$ are also collected in Table~\ref{tabku}.
We then obtain
\begin{equation}
 Z_{\cal O\,{\rm pert}}^{MF} =  1 - \frac{g_{TI}^2 C_F}{16 \pi^2} \left(1 -n_D
 - \frac{4 r}{\rho}\right) k_u^{TI} \,. 
\end{equation}
Let us now introduce constants $B_{\cal O}(\rho,C)$ by writing the
original one-loop renormalisation constants of Section 3 as 
\begin{equation}
Z_{\cal O} =  1 - \frac{C_F\, g^2}{16 \pi^2}
 \left[ \gamma_{\cal O} \log(a\mu) + B_{\cal O}(\rho,C) \right] \,,
 \label{BOdef} 
\end{equation}
where $\gamma_{\cal O}$ is the anomalous dimension of ${\cal O}$. We then
obtain tadpole improved renormalisation constants for $r=1$:
\begin{equation}
Z_{\cal O}^{TI} =  \frac{\rho \; u_0^{1-n_D} }{\rho -4 (1-u_0)}
 \left\{1 \!-\! \frac{g_{TI}^2 C_F}{16 \pi^2} \biggl[ \frac
 {\gamma_{\cal O}}{C_0^{TI}} \log(a\mu)
  + B_{\cal O}(\rho,C^{TI})
 - \!\biggl(\!1 -n_D-\frac{4}{\rho}\biggr)k_u^{TI}\biggr]\right\}\,,
\label{ZTI}
\end{equation}
where $B_{\cal O}(\rho,C^{TI})$ is the analogue of 
$B_{\cal O}(\rho,C)$, with $C_0$, $C_1$ and $C_2$ being
 replaced by $C_0^{TI}$, $C_1^{TI}$
and $C_2^{TI}$, respectively. In the following we shall use the abbreviation
\begin{equation}
B_{\cal O}^{TI} = B_{\cal O}(\rho,C^{TI})
 - \biggl(1 -n_D-\frac{4}{\rho}\,\biggr)k_u^{TI}\,,
\end{equation}
giving
\begin{equation}
Z_{\cal O}^{TI} = \frac{\rho \; u_0^{1-n_D} }{\rho -4(1- u_0)}
 \left\{1 - \frac{g_{TI}^2 C_F}{16 \pi^2} \left[
  \frac{\gamma_{\cal O}}{C^{TI}_0} \log(a\mu)
  + B_{\cal O}^{TI} \right]\right\}\,.
\end{equation}

So far we have only tadpole improved the gluon propagator in our calculation
of the perturbative factor on the r.h.s. of eq.~(\ref{zti}), but not the
fermion propagator. There is no need to do so for Wilson fermions. However,
already for clover fermions we have seen~\cite{Capitani:2000xi} that the
fermion propagator ought to be improved as well. If we want the rescaled
fermion propagator (see eq.~(\ref{finverse}))
\begin{equation}
\left[{\rm i} \pslash + \frac{1}{2}\, a\, p^2 \, \frac{u_0}{\rho
 -4r(1-u_0)}\right]^{-1}
\end{equation}
to have the same form as we change $u_0$, we must replace $\rho$ by
\begin{equation}
\rho^{TI} = \frac{\rho-4(1-u_0)}{u_0}
\label{rhoti}
\end{equation}
for $r=1$. An alternative derivation of $\rho^{TI}$, without
 expanding as a power series in $a$, is given in Appendix C. 
This defines `fully tadpole improved' renormalisation constants
\begin{equation}
Z_{\cal O}^{FTI} = \frac{\rho \; u_0^{1-n_D} }{\rho -4(1- u_0)}
 \left\{1 - \frac{g_{TI}^2 C_F}{16 \pi^2} \left[ \frac
 {\gamma_{\cal O} }{C_0^{TI}} \log(a\mu) + B_{\cal O}^{FTI} \right]\right\}\,,
\end{equation}
with
\begin{equation}
B_{\cal O}^{FTI} = B_{\cal O}(\rho^{TI},C^{TI})
 - \biggl(1 -n_D-\frac{4}{\rho^{TI}}\,\biggr)k_u^{TI}\,.
\end{equation}

Before we present numbers for $Z_{\cal O}^{TI}$ and $Z_{\cal O}^{FTI}$, let us
make a few general remarks concerning the choice of gauge action. Only if
$g_{TI}$ is a `good' expansion parameter, can we expect the
perturbative series to converge fast. It is generally accepted that
$g_\MS(\mu)$ is a good expansion parameter for appropriate choices of $\mu$. 
To one-loop order we have
\begin{equation}
\frac{1}{g^2_{\MS}(\mu)} - \frac{1}{g^2(a)}
 = 2 b_0 \left( \log \frac{\mu}{\Lambda_{\MS}}
- \log\frac{1}{a\Lambda_{\rm lat}}\right) = 2 b_0 \log(a\mu) + d_g 
+ N_f\, d_f \,, 
\end{equation}
where $b_0 = (11 - 2/3 N_f)/(4\pi)^2$, and $N_f$ is the number of flavours. (It
is appropriate to consider the case of general $N_f$ here.) The ratio of
$\Lambda$ parameters is thus given by
\begin{equation}
\frac{\Lambda_{\rm lat}}{\Lambda_{\overline{MS}}}
 = \exp \left(\frac{d_g + d_f}{2b_0}\right) \,.
\end{equation}
Upon inserting (\ref{gti}) and (\ref{exu}), we obtain 
\begin{equation}
\frac{1}{g^2_{\MS}(\mu)} - \frac{1}{g_{TI}^2(a)}
 = 2 b_0 \left( \log \frac{\mu}{\Lambda_{\MS}}
- \log\frac{1}{a\Lambda_{\rm lat}^{TI}}\right) = 2 b_0 \log(a\mu) + d_g 
+ N_f\,d_f + \frac{k_u^{TI}}{3\pi^2} \,,
\end{equation}
giving
\begin{equation}
\frac{\Lambda_{\rm lat}^{TI}}{\Lambda_{\overline{MS}}}
 = \exp \left(\frac{d_g + N_f\,d_f+k_u^{TI}/6\pi^2}{2b_0}\right) \,.
\end{equation}
The coefficient $d_g$ is known for some of our 
actions~\cite{dg}:
\begin{equation}
\begin{tabular}{c|c} \rule[-3mm]{0mm}{8mm}
Action & $d_g$ \\ \hline  &  \\[-1.25ex]
Plaquette & -0.4682 \\[0.7ex]
Symanzik & -0.2361 \\[0.7ex]
Iwasaki & \phantom{-}0.1053 \\[0.7ex]
DBW2 & \phantom{-}0.5317 
\end{tabular}
\label{dgtab}
\end{equation}
The coefficient $d_f$ is known for $\rho = 1.4$~\cite{Alexandrou:1999wr}:
\begin{equation}
d_f = -0.01449 \,.
\label{df}
\end{equation}
Unfortunately, $d_g$ is not known for the L\"uscher-Weisz action, nor is it
known for tadpole improved coefficients (\ref{c0ti}) and (\ref{cti}). We
expect the numbers for the L\"uscher-Weisz action to be close to the result for
the Symanzik action though. In Table~\ref{LamTab} we have computed the
ratios of $\Lambda$ parameters $\Lambda_{\rm lat}/\Lambda_\MS$ and
$\Lambda_{\rm lat}^{TI}/\Lambda_\MS$ for coefficients (\ref{dgtab}) and
(\ref{df}) and, to be consistent, with $k_u^{TI}$ replaced by $k_u$.
We see that tadpole improvement drives $\Lambda_{\rm lat}$ towards
$\Lambda_{\MS}$ for the Symanzik action, giving $g_{TI} \approx g_\MS$, so
that $g_{TI}$ appears to be a good expansion parameter. This is not the case
for the Iwasaki action, and even less so for the DBW2
action. We thus may conclude that the Symanzik action, and possibly the 
L\"uscher-Weisz action as well, is a suitable lattice gauge field action not
only from the non-perturbative perspective, but also from the perturbative
point of view, if supplemented with tadpole improvement.

\begin{table}[!t]
\begin{center}
  \begin{tabular}{|c|c|c|c|c|} \hline
  & \multicolumn{2}{|c|}{\rule[-3mm]{0mm}{8mm}$\Lambda_{\rm
 lat}/\Lambda_{\MS}$} &   
 \multicolumn{2}{|c|}{$\Lambda_{\rm lat}^{TI}/\Lambda_{\MS}$}\\
 \raisebox{1.5ex}[-1.5ex]{Action} & $N_f=0$ & $N_f=2$ & $N_f=0$ &
 $N_f=2$ \\ 
\hline   & & & & \\[-1.25ex]
Plaquette   & \phantom{0}0.0347 & \phantom{0}0.0172 & \phantom{0}0.380 &
 \phantom{0}0.262  \\ [0.7ex]
Symanzik & \phantom{0}0.184\phantom{0}  & \phantom{0}0.115\phantom{0} &
 \phantom{0}1.06\phantom{0} & \phantom{0}0.843  \\ [0.7ex] 
Iwasaki  & \phantom{0}2.13\phantom{00} & \phantom{0}1.86\phantom{00}  &
 \phantom{0}5.82\phantom{0} & \phantom{0}5.86\phantom{0}  \\ [0.7ex] 
DBW2     & 45.4\phantom{000}   & 60.7\phantom{000}   & 65.6\phantom{00}  &
 92.2\phantom{00}  \\ 
\hline
\end{tabular}
  \end{center}
  \caption{Ratios of $\Lambda$ parameters.}
  \label{LamTab}
\end{table}

\clearpage

\begin{table}[!htb]
\begin{center}
\begin{tabular}{lccccccc}
  \hline \rule[-3mm]{0mm}{10mm}
Action  & $\beta$   &
\multicolumn{1}{l}{$B_{S,P}$} & {$Z_{S,P}^\MS$}&
{$B_{S,P}^{TI}$}& {$Z_{S,P}^{TI,\MS}$}&
{$B_{S,P}^{FTI}$}& {$Z_{S,P}^{FTI,\MS}$}\\
  \hline
\rule[0mm]{0mm}{6mm}
  Symanzik      &  \phantom{0}8.45\phantom{0}    & $-13.876\phantom{0}$  & 1.139 &  &  &  &  \\ [0.7ex]
  TILW          &  \phantom{0}8.60\phantom{0}    & $-12.016\phantom{0}$  &1.164 & $-1.554$
  		&1.295 &$-3.329$ & 1.341\\ [0.7ex]
  TILW          &  \phantom{0}8.45\phantom{0}    & $-11.951\phantom{0}$  &1.168 & $-1.593$
  		&1.309 &$-3.421$ & 1.359\\ [0.7ex]
  TILW3          &  \phantom{0}8.30\phantom{0}   & $-11.862\phantom{0}$  &1.173 & $-1.644$
  		&1.325 &$-3.527$ & 1.380\\ [0.7ex]
  TILW          &  \phantom{0}8.20\phantom{0}    & $-11.806\phantom{0}$  &1.176 & $-1.677$
  		&1.337 &$-3.601$ & 1.395 \\ [0.7ex]
  TILW          &  \phantom{0}8.10\phantom{0}    & $-11.736\phantom{0}$  &1.180 & $-1.717$
  		&1.350 &$-3.683$ & 1.412 \\ [0.7ex]
  TILW          &  \phantom{0}8.00\phantom{0}    & $-11.648\phantom{0}$  &1.184 & $-1.766$
  		&1.366 &$-3.778$ & 1.432 \\ [0.7ex]
  Iwsasaki       &  \phantom{0}9.485   & $-9.851$  & 1.192 & $-2.527$  &1.334
  		&$-3.851$ & 1.381 \\ [0.7ex]
  Iwsasaki       &  \phantom{0}8.026   & $-9.851$  & 1.227 & $-2.610$  &1.481
  		&$-4.379$ & 1.573 \\ [0.7ex]
  DBW2          &  12.745  & $-7.270$  & 1.354 & $-4.095$  & 1.506
  		&$-4.543$ & 1.541 \\ [0.7ex]
  DBW2          &  10.662  & $-7.270$  & 1.423 & $-4.122$  & 1.681
  		&$-4.666$ & 1.739 \\ [0.7ex]
  \hline
  \end{tabular}
  \end{center}
  \caption{The constants $B_{S,P}$ and $Z_{S,P}^\MS(a\mu=1)$ for various
  levels of improvement.}  
  \label{ZSTab}
\end{table}
\vspace*{0.5cm}

\begin{table}[!htb]
\begin{center}
 \begin{tabular}{lccccccc}
  \hline \rule[-3mm]{0mm}{10mm}
Action  & $\beta$   &
\multicolumn{1}{l}{$B_{V,A}$} & {$Z_{V,A}^\MS$}&
{$B_{V,A}^{TI}$}& {$Z_{V,A}^{TI,\MS}$}&
{$B_{V,A}^{FTI}$}& {$Z_{V,A}^{FTI,\MS}$}\\
  \hline
\rule[0mm]{0mm}{6mm}
  Symanzik      &  \phantom{0}8.45\phantom{0}    & $-13.397\phantom{0}$  & 1.134 &  &  &  &  \\ [0.7ex]
  TILW          &  \phantom{0}8.60\phantom{0}    & $-11.180\phantom{0}$  &1.153 & $-0.099$
  		&1.258 &$-1.342$ & 1.290\\ [0.7ex]
  TILW          &  \phantom{0}8.45\phantom{0}    & $-11.100\phantom{0}$  &1.156 & $-0.101$
  		&1.268 &$-1.384$ & 1.303\\ [0.7ex]
  TILW3          &  \phantom{0}8.30\phantom{0}   & $-10.990\phantom{0}$  &1.160 & $-0.105$
  		&1.280 &$-1.431$ & 1.319\\ [0.7ex]
  TILW          &  \phantom{0}8.20\phantom{0}    & $-10.922\phantom{0}$  &1.163 & $-0.106$
  		&1.289 &$-1.464$ & 1.330 \\ [0.7ex]
  TILW          &  \phantom{0}8.10\phantom{0}    & $-10.835\phantom{0}$  &1.166 & $-0.109$
  		&1.299 &$-1.410$ & 1.343 \\ [0.7ex]
  TILW          &  \phantom{0}8.00\phantom{0}    & $-10.727\phantom{0}$  &1.169 & $-0.112$
  		&1.311 &$-1.539$ & 1.358 \\ [0.7ex]
  Iwsasaki       &  \phantom{0}9.485   & $-8.362$  & 1.163 & $-0.271$  &1.252
  		&$-1.196$ & 1.285 \\ [0.7ex]
  Iwsasaki       &  \phantom{0}8.026   & $-8.362$  & 1.193 & $-0.204$  &1.356
  		&$-1.469$ & 1.422 \\ [0.7ex]
  DBW2          &  12.745  & $-3.538$  & 1.172 & $-0.277$  & 1.204
  		&$-0.582$ & 1.228 \\ [0.7ex]
  DBW2          &  10.662  & $-3.538$  & 1.206 & $-0.230$  & 1.264
  		&$-0.607$ & 1.304 \\ [0.7ex]
  \hline
  \end{tabular}
  \end{center}
  \caption{The same as Table~\ref{ZSTab}, but for $B_{V,A}$ and
    $Z_{V,A}^\MS(a\mu=1)$.}
\label{ZVTab}
\end{table}

\clearpage
\begin{table}[!htb]
\begin{center}
 \begin{tabular}{lccccccc}
  \hline \rule[-3mm]{0mm}{10mm}
Action  & $\beta$   &
\multicolumn{1}{l}{$B_{T}$} & {$Z_{T}^\MS$}&
{$B_{T}^{TI}$}& {$Z_{T}^{TI,\MS}$}&
{$B_{T}^{FTI}$}& {$Z_{T}^{FTI,\MS}$}\\
  \hline
\rule[0mm]{0mm}{6mm}
  Symanzik      &  \phantom{0}8.45\phantom{0}    & $-14.861\phantom{0}$  & 1.148 &  &  &  &  \\ [0.7ex]
  TILW          &  \phantom{0}8.60\phantom{0}    & $-12.750\phantom{0}$  &1.174 & $-1.100$
  		&1.283 &$-2.128$ & 1.310\\ [0.7ex]
  TILW          &  \phantom{0}8.45\phantom{0}    & $-12.675\phantom{0}$  &1.178 & $-1.082$
  		&1.295 &$-2.146$ & 1.324\\ [0.7ex]
  TILW3          &  \phantom{0}8.30\phantom{0}   & $-12.571\phantom{0}$  &1.183 & $-1.062$
  		&1.308 &$-2.163$ & 1.340\\ [0.7ex]
  TILW          &  \phantom{0}8.20\phantom{0}    & $-12.507\phantom{0}$  &1.186 & $-1.048$
  		&1.318 &$-2.177$ & 1.352 \\ [0.7ex]
  TILW          &  \phantom{0}8.10\phantom{0}    & $-12.424\phantom{0}$  &1.190 & $-1.032$
  		&1.328 &$-2.189$ & 1.365 \\ [0.7ex]
  TILW          &  \phantom{0}8.00\phantom{0}    & $-12.322\phantom{0}$  &1.194 & $-1.012$
  		&1.341 &$-2.202$ & 1.380 \\ [0.7ex]
  Iwsasaki       &  \phantom{0}9.485   & $-10.130$  & 1.197 & $-1.029$  &1.279
  		&$-1.773$ & 1.306 \\ [0.7ex]
  Iwsasaki       &  \phantom{0}8.026   & $-10.130$  & 1.233 & $-0.806$  &1.388
  		&$-1.851$ & 1.442 \\ [0.7ex]
  DBW2          &  12.745  & \phantom{0}$-6.021$  & 1.293 & $-0.364$  & 1.210
  		&$-0.581$ & 1.228 \\ [0.7ex]
  DBW2          &  10.662  & \phantom{0}$-6.021$  & 1.351 & $-0.164$  & 1.257
  		&$-0.446$ & 1.287 \\ [0.7ex]
  \hline
  \end{tabular}
  \end{center}
   \caption{The same as Table~\ref{ZSTab}, but for $B_{T}$ and
    $Z_{T}^\MS(a\mu=1)$.}
\label{ZTTab}
\end{table}

In Tables \ref{ZSTab},~\ref{ZVTab} and \ref{ZTTab} we compare tadpole improved
and unimproved renormalisation constants for $\rho=1.4$ and some selected
values of $\beta$,
which are widely being used in Monte Carlo simulations. The DBW2 couplings
$\beta = 10.662$ and 12.745 correspond to $\beta = 0.87$ and 1.04,
respectively, in the notation of~\cite{Aoki:2002vt}. The `measured' and
perturbative plaquette values are taken from Table~\ref{tabku}. The tadpole
improved renormalisation constants given earlier in~\cite{Galletly:2003vf}
referred to 
unimproved coefficients $c_i$ and unimproved parameter $\rho$.

\begin{figure}[b!]
\vspace*{0.75cm}
\begin{center}
\epsfig{file=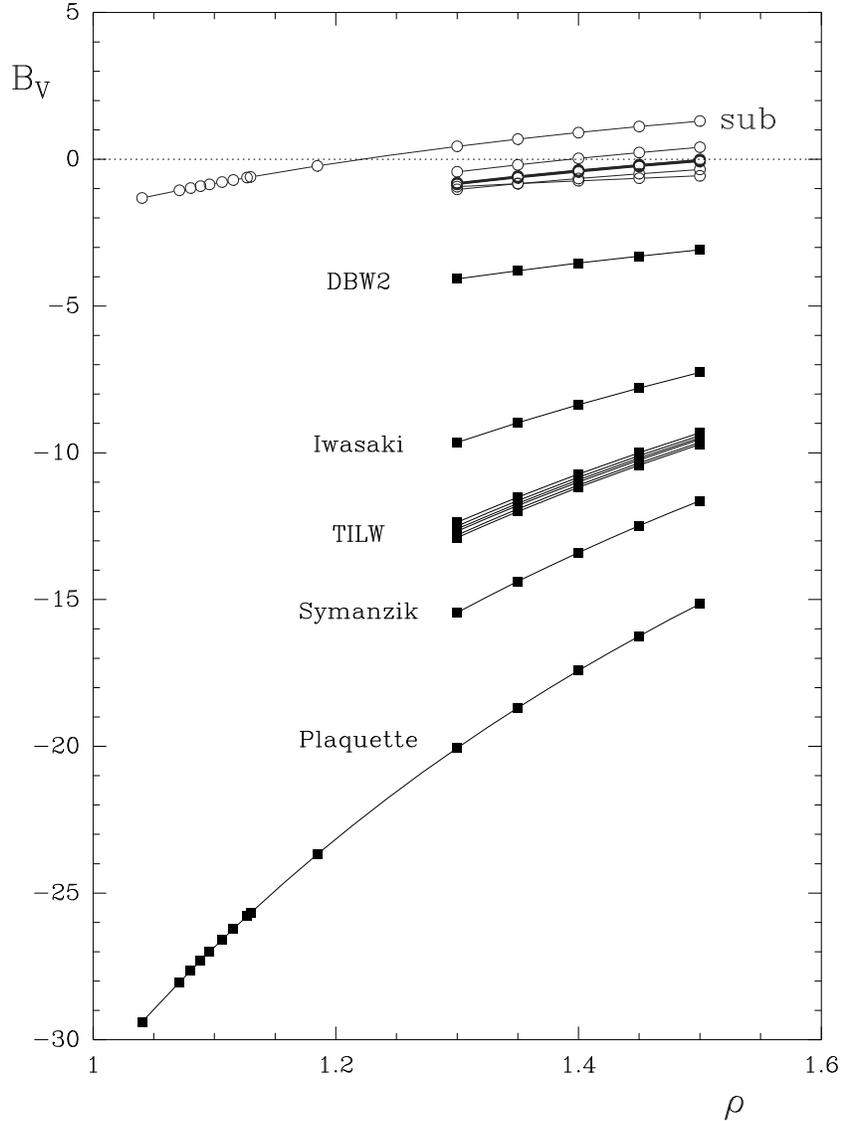,width=11cm}
\end{center}
\caption{A test of the mean field approximation. We compare the 
 perturbative $B_V$ (filled squares), as defined in (\ref{BOdef}), 
 with the subtracted $B_V^{sub}$ (open circles)
 defined in (\ref{Bsub}). 
 The subtracted coefficients are much smaller, showing that the
 mean field approximation is very good at the one-loop level. }
  \label{Btad}
\end{figure}

\begin{figure}[t!]
\begin{center}
\epsfig{file=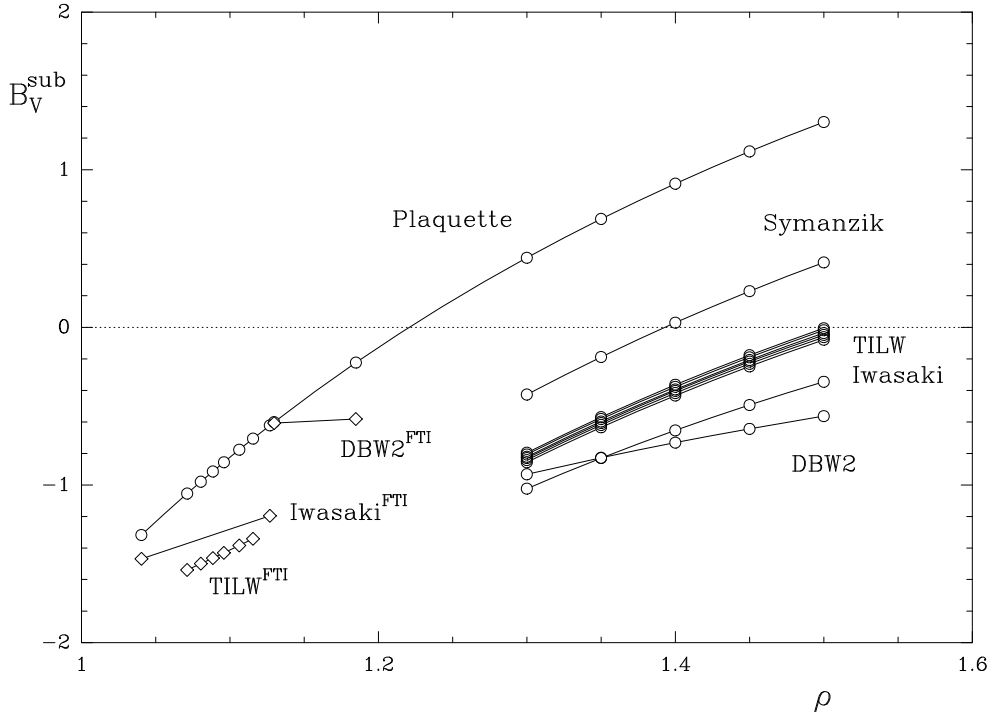,angle=270,width=13cm}
\end{center}
\caption{The subtracted points from Fig.~\ref{Btad} shown in 
 more detail. The open circles show results calculated 
 with unimproved gauge actions, the open diamonds with gauge
 actions modified according to eqs.~(\ref{c0ti}), (\ref{cti}) and 
(\ref{rhoti} ). All actions considered give very similar results. } 
  \label{BTI}
\end{figure}

    How good is the mean field approximation and tadpole improvement? 
 The ultimate test is to compare with non-perturbative determinations 
 of $Z$, which at present we can only do in a single case (to be discussed
 later). However, there is also a powerful internal test, which we can 
 make with our perturbative coefficients.  If the mean field $Z^{MF}$
 of eqs.~(\ref{Zpsimf}) and (\ref{ZOmf}) is a good approximation,
 then we should always find that the one-loop coefficient of
 $Z^{MF}$ is very close to the one-loop coefficient of $Z_\psi$ or
 $Z_O$, i.e. that 
 \begin{equation}
 B_O^{sub}(\rho, C) \equiv 
 B_O(\rho,C) - \left(1-n_D -\frac{4}{\rho}\right)\, k_u
 \label{Bsub}
 \end{equation}
 is small, whatever gauge action or $\rho$ we use.
 We test this in Fig.~\ref{Btad}, where we compare the original
 values of $B_V$ with the subtracted ones defined in (\ref{Bsub}).
 The original coefficients $B_V$ have large negative values, and depend
 strongly 
 on the choice of gauge action and on the value of $\rho$. 
 The subtracted coefficients $B_V^{sub}$ are much closer to zero, 
 and depend only weakly on gauge action or $\rho$. So we see that
 the mean field approximation of
 eqs.~(\ref{Zpsimf}) and (\ref{ZOmf}) is very good at the one-loop level.
 If we had naively used the Wilson fermion result~(\ref{zmfw}),
 this test would have failed completely, not even having the right sign.  
 In Fig.~\ref{BTI} we show the subtracted coefficients in more detail.
 We can see that they all lie in a fairly narrow band, with points calculated
 from the original actions (solid points) and the tadpole improved actions
 (open points) in agreement. This shows how well the dependence of
 $Z$ on gauge action and $\rho$ is described by the mean field approximation, 
 leaving only a small residue to be described by perturbation theory. 

As far as we can tell from the single non-perturbative result we have found
for the TILW action at $\beta=8.45$~\cite{Galletly:2003vf}), corresponding to
a lattice spacing of $a \approx 0.1$ fm, tadpole
improvement drives the perturbative number closer to the non-perturbative
value. But the discrepancy is still of the order of 10\%.

An alternative improvement scheme has been proposed in~\cite{Kronfeld:2003sd},
in which $Z_\mathcal{O}^{MF}$ is 
replaced by the non-perturbatively computed renormalisation constant of the
local vector current $Z_V^{NP}$:
\begin{equation}
Z_{\cal O}^{VI} = Z_V^{NP} \left(\frac{Z_{\cal O}}{Z_V}\right)_{\rm pert}\,.
\end{equation}
This gives
\begin{equation}
\begin{split}
Z_{\cal O}^{VI} &= Z_V^{NP} \left\{1-\frac{C_F g_{TI}^2}{16\pi^2}
\left[\frac{\gamma_{\cal O}}{C_0^{TI}} \log(a\mu) + B_{\cal O}^{FTI} -
  B_V^{FTI}\right]\right\} \\ 
  &\equiv
   Z_V^{NP} \left\{1-\frac{C_F g^2_{TI}}{16\pi^2}
\left[\frac{\gamma_{\cal O}}{C_0^{TI}} \log (a\mu) + B_{\cal
    O}^{VI}\right]\right\}\,. 
\end{split}
\label{kronfeld}
\end{equation}
For the TILW action at $\beta=8.45$ and $\rho=1.4$ we found
$Z_V^{NP} (= Z_A^{NP}) = 1.416$~\cite{Galletly:2003vf}. We used this
number to 
compute $Z_{\cal O}^{VI}$ in Table~\ref{ZVITab}.
This method is not applicable to operators with $n_D > 0$ covariant
derivatives. In that case one would have to replace the local vector current
by an operator with an equal number (i.e.~$n_D$) of covariant derivatives.

\begin{table}[t!]
\begin{center}
  \begin{tabular}{|c|c|c|}
  \hline \rule[-3mm]{0mm}{10mm}
$Z_{S,P}^{VI,\MS}$ & $Z_{V,A}^{NP}$ & $Z_{T}^{VI,\MS}$\\
  \hline
\rule[0mm]{0mm}{6mm}
  1.478 & 1.416 & 1.439 \\ [0.7ex]
  \hline
  \end{tabular}
  \end{center}
   \caption{The renormalisation constants $Z_{S,P}^{VI,\MS}(a\mu=1)$ and
     $Z_{T}^{VI,\MS}(a\mu=1)$ for the TILW action at $\beta=8.45$. The value
     of $Z_{V,A}^{NP}$ was used as input.} 
\label{ZVITab}
\end{table}

\section{Summary}

We have computed the renormalisation constants of local quark bilinear
operators for overlap fermions and improved gauge field actions with up to six
links. The computations have been performed in general covariant gauge using
the symbolic language {\sl Mathematica}. This gave us complete control over 
the Lorentz and spin structure, the cancellation of infrared divergences, as
well as the cancellation of $1/a$ singularities. The price to pay is high.
The extension to improved gauge field actions blew up the calculation 
tremendously. In some instances we had to deal with $O(10^4)$ terms.
Based on generalised lattice Ward identities we were able to show
analytically that the gauge dependent part of the self-energy and the
amputated Green functions does not depend on the choice of lattice fermions.

In the limit $c_0=1$, $c_1=c_2=c_3=0$ our results agree with previous
calculations employing the plaquette
action~\cite{Alexandrou:2000kj,Capitani:2000wi}. Comparing overlap with
Wilson fermions, we notice that the one-loop corrections have opposite sign.
This is mainly caused by changes in the quark self-energy, and can be
 understood through a mean field calculation. 

We have formulated mean field (tadpole) improvement for overlap fermions. For
the Symanzik action the boosted coupling $g_{TI}$ turns out to be close to 
$g_\MS$, which makes $g_{TI}$ a good expansion parameter. We thus may expect 
that the perturbative series converges rapidly. For the Iwasaki action, and in 
particular for the DBW2 action, this appears not to be true. In this case
boosted perturbation theory might even worsen the situation.

We have presented results for a variety of parameters and couplings, which
cover most of the parameter values used in recent Monte Carlo simulations. We
are happy to supply numbers for different choices of parameters on request. 
Details of our results for operators with covariant
derivatives~\cite{Galletly:2003vf}  will be given elsewhere.  

\section*{Acknowledgements}

This work is supported by DFG under contract FOR 465 (Forschergruppe
Gitter-Hadronen-Ph\"{a}nomenologie), which we gratefully acknowledge. We thank
Tom Blum for providing us with the plaquette values for the DBW2 action.
The authors would like to thank Haris Panagopoulos for pointing out the
misprints in Appendix A.

\section*{Appendix A}

We set $a=1$ throughout this Appendix. Let us first consider the gluon
propagator.
We introduce the abbreviation
\begin{equation}
\hat{k}^n=\sum_{\mu=1}^4 \hat{k}_\mu^n\,, \quad n=2,4,\cdots\,.
\end{equation}
The coefficients $D_n$ and $D_{m,n}$ in eq.~(\ref{DeltaD})
can be written in the form
\begin{equation}
D_n = \frac{d_n}{D\,(\hat{k}^2)^2}, \quad D_{m,n} = \frac{d_{m,n}}{D\,(\hat{k}^2)^2} \,,
\end{equation}
where $d_{m,n}=d_{n,m}$. For $D$, in the case $C_0=1$, we find
\begin{equation}
\begin{split}
D &=  (1 - C_1\, \hat k^2)^3\, (\hat k^2)^2- C_2 \, (1 - C_1\, \hat
 k^2)^2\,\hat k^2 \, \Bigl[(\hat k^2)^2 + 2\, \hat k^4\Bigr]  \\[0.7ex]
 &\phantom{=} + \frac{1}{2} \,C_2^2 \, (1 - C_1\, \hat k^2) \, \Bigl[(\hat
 k^2)^4 + 
 (\hat k^2)^2\, \hat k^4 + 2\, (\hat k^4)^2 + 2 \,\hat k^2\, \hat
 k^6\Bigr]  \\[0.7ex]
 &\phantom{=} + \frac{1}{6}\,C_2^3 \,\Bigl[ 24 \,\hat k^{10} - 7 \, (\hat
 k^2)^3\, 
 \hat k^4 +  15\, \hat k^2\, (\hat k^4)^2  + 12\, (\hat k^2)^2\,
 \hat k^6 \\[0.7ex] &\phantom{=} - 26\, \hat k^4\, \hat k^6 - 24\, \hat k^2\,
 \hat k^8 
 \Bigr] \,, 
\end{split}
\end{equation}
and for $d_n$ and $d_{m,n}$ we obtain
\begin{equation}
\begin{split}
d_{0\phantom{,0}} &= - D \, \hat k^2 + (1 - C_1\,\hat k^2)^2 \,(\hat
k^2)^3 -C_2\, (1 - C_1\,\hat k^2) \,
\Bigl[ 7\, (\hat k^2)^2 \hat k^4 \\&\phantom{=} - 3 \, (\hat k^4)^2 - 8 \,
\hat k^2  \, \hat k^6 + 6 \, \hat k^8 \Bigr]  %\\ 
+ \frac{1}{2}\, C_2^2 \,\Bigl[30\, \hat k^{10} + 9\,(\hat
k^2)^2\, \hat k^6 \\&\phantom{=} - 25\,\hat k^4 \, \hat k^6 +
  12\, \hat k^2\,((\hat k^4)^2 - 2\, \hat k^8)\Bigr] \, ,
\\[0.7ex]
d_{1\phantom{,0}} &= C_2 \,(1 - C_1\,\hat k^2) \,  \Bigl[-2 \, (\hat
k^2)^3 + 9\, \hat k^2 \, \hat k^4- 6\, \hat k^6 \Bigr]\\
&\phantom{=} + \,C_2^2 \, \Bigl[ 4\, (\hat k^2)^2 \hat k^4 - 7 \, (
k^4)^2 - 12 \, \hat k^2 \, \hat k^6 + 14 \, \hat k^8 \Bigr] \,, \\[0.7ex]
d_{2\phantom{,0}} &= 6\, C_2 \,(1 - C_1\,\hat k^2) \, \Bigl[(\hat k^2)^2
- \hat k^4 \Bigr]
+ \frac{1}{2}C_2^2 \, \Bigl[ 3\, (\hat k^2)^3
- 29 \, \hat k^2 \, \hat k^4 \\&\phantom{=} + 28 \, \hat k^6 \Bigr] \,,
\\[0.7ex] 
d_{3\phantom{,0}} &= - 6\, C_2 \,(1 - C_1\,\hat k^2) \,\hat k^2
- 2\, C_2^2 \, \Bigl[4 (\hat k^2)^2 - 7\,\hat k^4 \Bigr]\,,  \\[0.7ex]
d_{4\phantom{,0}} &= 15\,C_2^2\,\hat k^2  \,,\\[0.7ex]
d_{0,0} &= D -(1 - C_1 \,\hat k^2)^2 (\hat k^2)^2 +
C_2 \, (1 - C_1\,\hat k^2)\,
\Bigl[5 \, \hat k^2 \, \hat k^4  - 2\,\hat k^6 \Bigr]\\ 
&\phantom{=} + C_2^2\, \Bigl[-3  (\hat k^4)^2 - 3\,\hat k^2 \, \hat k^6
+ 4 \, \hat k^8 \Bigr] \,,\\[0.7ex]
d_{1,1} &= - 6\, C_2 \,(1 - C_1\,\hat k^2)\, \hat k^2
- C_2^2 \, \Bigl[(\hat k^2)^2 - 7\,\hat k^4 \Bigr] \,,  \\[0.7ex]
d_{2,2} &=  d_{1,3} \, \,= -14\,C_2^2 \,, \\[0.7ex]
d_{0,1} &=
C_2 \,(1 - C_1\,\hat k^2) \,  \Bigl[2 \, (\hat k^2)^2
- 3\, \hat k^4 \Bigr]
- C_2^2 \, \Bigl[ 3\, \hat k^2 \, \hat k^4 - 5 \, \hat k^6 \Bigr]\,,
\\[0.7ex] 
d_{0,2} &=
-6\, C_2 \,(1 - C_1\,\hat k^2)\,\hat k^2  -
\frac{3}{2}\,  C_2^2 \, \Bigl[ (\hat k^2)^2 - 5 \, \hat k^4 \Bigr]\,,
\\[0.7ex] 
d_{0,3} &= 6\, C_2 \,(1 - C_1\,\hat k^2)\, + 8 \,C_2^2 \, \hat k^2\,,
\\[0.7ex] 
d_{0,4} &= -15\,C_2^2\,, \\[0.7ex]
d_{1,2} &= 6\, C_2 \,(1 - C_1\,\hat k^2)\, +
7 \,C_2^2 \, \hat k^2  \,.
\end{split}
\end{equation}
(The coefficient $d_{m,n}$ should not be confused with $d_{\mu\nu}$ of
eq.~(\ref{abbrev})). Note that $D_n$ and $D_{m,n}$ do not depend on the choice
of (covariant) gauge, i.e. on $\xi$. Both $D_n$ and $D_{m,n}$ vanish in the
limit $C_1 = C_2 = 0$. 

Let us now turn to the lattice Feynman rules. We omit the colour factors and
the gauge coupling here.
We introduce the following abbreviations:
\bea
\omega(p) \;= \; \omega(-p) &=& \sqrt{\sum_\mu \sin^2 p_\mu + b(p)^2 }\,, \quad
b(p)=b(-p)= \frac{r}{2}\, {\hat p}^2 - \rho \,,\\
\omega(p_2,p_1) &=& \sqrt{\sum_\mu \sin p_{2\,\mu}\sin p_{1\,\mu} +
  b(p_2)b(p_1)
}\,, \\
X_0(p) &=& {\rm i} \sum_\mu \gamma_\mu \sin p_\mu + b(p)\,, \quad
X_0^\dagger(p) X_0(p)= \omega(p)^2\,. \label{X0}
\eea
The Wilson quark propagator and its inverse are of the form
\bea
S_W(p)=\frac{\displaystyle -{\rm i} \sum_\mu \gamma_\mu \sin p_\mu +
  \frac{r}{2}\, \hat p^2} {\displaystyle \sum_\mu \sin^2 p_\mu +
  \left(\frac{r}{2}\, \hat p^2\right)^2 }
%= \frac{X_0^\dagger(p)+\rho}{(\omega(p))^2 - (b(p))^2 + (b(p)+\rho)^2}
\,, \quad
S_W^{-1}(p)={\rm i} \sum_\mu \gamma_\mu \sin p_\mu + \frac{r}{2}\,
\hat p^2= X_0(p)+\rho \,. 
\eea
The Wilson  quark-quark-gluon (qqg) and quark-quark-gluon-gluon (qqgg)
vertices are given by 
\bea
V_{1\,\mu}^W(p_2,p_1)&=& - {\rm i}\, \gamma_\mu \cos \frac{(p_2+p_1)_\mu}{2} -
r \sin \frac{(p_2+p_1)_\mu}{2}\, , \label{qqgwilson} \\
V_{2\,\mu\nu}^W(p_2,p_1)&=& - \frac{1}{2} \delta_{\mu\nu} \left[ -{\rm i}\,
\gamma_\mu \sin \frac{(p_2+p_1)_\mu}{2} + r \cos \frac{(p_2+p_1)_\mu}{2}
\right]\,,
\eea
with incoming and outgoing quark momenta being denoted by $p_1$ and $p_2$,
respectively. Note that
\bea
X_0(p)+\rho = S_W^{-1}(p) = -\sum_\mu \hat p_\mu V_{1\,\mu}^W(p,0)\equiv - \hat
p\, V_1^W(p,0) \,. 
\label{short}
\eea
The overlap quark propagator and its inverse are of the form
\be
S_N(p)=  \frac{X_0^\dagger(p) +\omega(p)}{2 \rho \left[\omega(p)+b(p)\right]}
\,, \quad
S_N^{-1}(p)=\rho \left[ \frac{X_0(p)}{\omega(p)} +1 \right] \,.
\ee
For the overlap qqg vertex we obtain
\be
V_{1\,\mu}^N(p_2,p_1) = \frac{\rho}{\omega(p_2)+\omega(p_1)}
\left[ V_{1\,\mu}^W(p_2,p_1)-\frac{X_0(p_2)}{\omega(p_2)}
  V_{1\,\mu}^{W\,\dagger}(p_2,p_1) \frac{X_0(p_1)}{\omega(p_1)}\right]\,,
\label{qqgover}
\ee
and for the qqgg vertex we derive
\bea
V_{2\,\mu\nu}^N(p_2,p_1,k_1,k_2) =
V_{{21}\,\mu\nu}^N(p_2,p_1)+V_{{22}\,\mu\nu}^N(p_2,p_1,k_1,k_2) \,,
\label{qqggover}
\eea
with
\bea
\!\!\!\!\!\!V_{{21}\,\mu\nu}^N(p_2,p_1) &\!=&\!
 \frac{\rho}{\omega(p_2)+\omega(p_1)}
\biggl[ V_{2\,\mu\nu}^W(p_2,p_1)-\frac{X_0(p_2)}{\omega(p_2)}
V_{2\,\mu\nu}^{W\,\dagger}(p_2,p_1) \frac{X_0(p_1)}{\omega(p_1)}
\biggr]\, ,\label{v2over1} \\[0.7ex]
\label{v2over2}
\!\!\!\!\!\!V_{{22}\,\mu\nu}^N(p_2,p_1,k_1,k_2) &\!=&\!
\frac{\rho}{2(\omega(p_2)+\omega(p_1))} 
\biggl[ W_{\mu\nu}(p_2,p_1,k_1) + W_{\nu\mu}(p_2,p_1,k_2)
\biggr]
\,,
\eea
where $k_1$ and $k_2$ are the gluon momenta with $p_1+k_1+k_2=p_2$, and the
tensor $W_{\mu\nu}$ is defined by
\be
\label{W}
\begin{split}
W_{\mu\nu}(p_2,p_1,k) &=\frac{1}{\omega(p_2)+\omega(p_1+k)}\,
\frac{1}{\omega(p_1+k)+\omega(p_1)} \,\biggl[ X_0(p_2)\,
V_{1\,\mu}^{W\,\dagger}(p_2,p_1+k) \\[1ex]
&\phantom{=} \times V_{1\,\nu}^W(p_1+k,p_1)
+ V_{1\,\mu}^W(p_2,p_1+k)\, X_0^\dagger(p_1+k)\, V_{1\,\nu}^W(p_1+k,p_1)\\[1ex]
&\phantom{=} + V_{1\,\mu}^W(p_2,p_1+k)\, V_{1\,\nu}^{W\,\dagger}(p_1+k,p_1)\,
X_0(p_1) \\[1ex] &\phantom{=}
-\frac{\omega(p_2)+\omega(p_1+k)+\omega(p_1)}
{\omega(p_2)\,\omega(p_1+k)\,\omega(p_1)}X_0(p_2)\,
V_{1\,\mu}^{W\,\dagger}(p_2,p_1+k)\, X_0(p_1+k)\,\\[1ex] &\phantom{=} \times
V_{1\,\nu}^{W\,\dagger}(p_1+k,p_1)\, X_0(p_1)
\biggr] \,.
\end{split}
\ee

In Appendix B we need explicit expressions for the vertices contracted with
the gluon momentum $k=p_2-p_1$ and $k=k_1=-k_2$, respectively:
\be
\begin{split}
\hat k\, V_1^{W,N}(p_2,p_1) &= \sum_\mu \hat k_\mu V_{1\,\mu}^{W,N}(p_2,p_1)\,,
\\ 
\hat k\, V_{2}^{W,N}(p,p,k,-k)\, \hat k &= \sum_{\mu\nu} \hat k_\mu
V_{2\,\mu\nu}^{W,N}(p,p,k,-k)\, \hat k_\nu \,.
\end{split}
\ee
Using the short-hand notation
\be
\omega = \omega(k)\,, \quad b = b(k) \,, \quad X_0 = X_0(k)\,,\quad
\hat k\, V_1^{W,N} = \hat k\, V_1^{W,N}(k,0) = \hat k\, V_1^{W,N}(0,k)\,,
\label{ob}
\ee
we obtain 
\be
\label{repl}
\hat k\, V_1^{W\,\dagger} \, \hat k\, V_{1}^W = \omega^2 + 2 \rho \, b + \rho^2
\,, \quad
\hat k\, V_{1}^W\,  X_0^\dagger \, \hat k\, V_{1}^W =
2 \omega^2 \rho + \omega^2 X_0 + \rho^2 X_0^\dagger 
\ee
and
\be
\hat k\,  V_{1}^N
= \frac{\rho}{\omega} \left( \hat k\, V_1^W + \rho -\omega \right)
\,.
\label{kV1O}
\ee
Differentiating $X_0^\dagger X_0=\omega^2$ with respect to $k_\mu$,
and making use of $V_{1\,\mu}^W(k,k)=- \partial X_0 / \partial k_\mu$ we find
\be
V_{1\,\mu}^N(k,k) =\frac{\rho}{\omega^3} \biggl[
\omega^2 V_{1\,\mu}^W(k,k)+ (\cos k_\mu + r b )\, \sin k_\mu \, X_0 \biggr]\,.
\label{V1Oalpha}
\ee
From
\be
\begin{split}
&\hat k\, V_1^W(p+k,p) = X_0(p)-X_0(p+k)\,,\\[0.7ex]
&X_0(p)\,\hat k\, V_1^{W\,\dagger}(p,p+k) + \hat k\, V_1^W(p,p+k)\,
X_0(p+k) = \omega(p)^2-\omega(p+k)^2\, , \\[0.7ex]
&X_0(p) X_0^\dagger(p+k) + X_0(p+k) X_0^\dagger(p) = 2\, \omega(p,p+k)^2
\end{split}
\ee
we finally obtain
\be
\begin{split}
\hat k\, V_{22}^N(p,p,k,-k)\,\hat k &= \frac{\rho}{\omega(p)}
\Biggl[\frac{\omega(p)^2- \omega(p,p+k)^2}{\omega(p)^2} X_0(p)\\[0.7ex]
&\phantom{=} + \frac{\omega(p+k)-\omega(p)}{\omega(p+k)} X_0(p+k) \Biggr] \,.
\end{split}
\label{kV22knew}
\ee

\section*{Appendix B}

\noindent
{\bf Gauge dependence in position space}

\noindent
{\it General argument}

We have seen in the main text that the gauge dependent terms in the Green
functions are independent of the gauge action and of the fermion action: they
do not depend on the coefficients $c_1, c_2, c_3$ and on $\rho$, and are the
same for overlap fermions and clover fermions~\cite{Capitani:2000xi}. The
numbers are, however, different on the lattice and in the continuum, so this
is not simply due to an ultraviolet convergent integral, where the lattice
regularisation plays no role. The independence from the gauge action is fairly
easy to explain, but the independence of the fermion action is a remarkable
observation, and is something which we ought to understand. 

In this subsection we give a brief explanation of this phenomenon in
configuration space, while in the next subsection we give a more formal
argument using the Feynman rules.

At the one-loop level we can split the functional integral over the gauge
fields into an integral over the (physical) transverse gauge fields
and the (unphysical) longitudinal gauge fields. The integral over the
longitudinal fields is a simple Gaussian integral, given by the gauge fixing
term added to the action. The contribution of the longitudinal modes is a
universal factor multiplying the Landau gauge Green function
\begin{equation}
G(x,g^2,\xi) = G(x,g^2,\xi=1)
\left[ 1 - (1-\xi)\, g^2 C_F
\int \frac{d^4 k} {(2 \pi)^4} \, \frac{ 1 -\cos kx}
{ (\hat k^2)^2 } + \cdots \right]\,,
\label{startpoint}
\end{equation}
where $G$ can be any quark Green function, e.g. the quark propagator or any of
the three-point functions we consider in this paper. See Appendix A
of~Ref.~\cite{QED}, where this is proved to all orders for non-compact QED. It
is not clear whether this result will hold to all orders for a non-Abelian
group, but it is certainly true at the one-loop level.

Equation~(\ref{startpoint}) gives us an exact expression for the $O(g^2)$
gauge dependent term in configuration space 
\begin{equation}
\label{exactpos}
G^{\rm gauge}(x) = g^2 C_F\,  \xi \, G^{\rm tree}(x)
\int \frac{d^4 k} {(2 \pi)^4} \, \frac{ 1 -\cos kx}
{ (\hat k^2)^2 } \,. 
\end{equation}
When we Fourier transform this expression to find the gauge term in momentum
space, the integral is dominated by $x^2 \sim 1/p^2$. Since we are interested
in $a^2 p^2 \ll 1$, we have to consider what happens to eq.~(\ref{exactpos})
when $x^2 \gg a^2$. Although the tree-level Green functions do depend on the
fermion action when $x^2 \sim a^2$, at large distances all fermion actions
agree: they simply tend to the continuum value $G^{\rm tree}_{\rm cont}(x)$.
Since this is the region that dominates the Fourier transform, we see
immediately that the gauge term will be the same for all fermion actions (or
to be more precise, that differences in the gauge term due to differing
choices of fermion are suppressed by powers of $a$). This completes our
argument for the universality of the gauge dependent terms.

Note that although the gauge dependent terms are independent of both gauge and
fermion action, they would change if we used a different lattice
discretisation of the gauge fixing term $(\partial_\mu A_\mu)^2$, as advocated
in~\cite{Bonnet:2000kw}. 

\noindent
{\it Calculating the gauge dependent contributions}

Equation~(\ref{exactpos}) not only shows that the gauge dependent terms are
universal, it also allows us to calculate them. At large distances ($x^2 \gg
a^2$) the integral in eq.~(\ref{exactpos}) takes the value 
\begin{equation}
\int \frac{d^4 k} {(2 \pi)^4} \, \frac{ 1 -\cos kx}
{ (\hat k^2)^2 } \quad \longrightarrow \quad
\frac{1}{16 \pi^2} \Big(F_0 + \gamma_E + \log \frac{x^2}{4 a^2} \Big) \,.
\end{equation}
In position space the massless quark propagator $S^{\rm tree}_{\rm cont}(x)$
and the three-point function $G^{\rm tree}_{\rm cont}(x)$ are
\be
\begin{split}
S^{\rm tree}_{\rm cont}(x) &= \frac{1}{2 \pi^2} \frac{\xslash}{(x^2)^2}\,,
\\
G^{\rm tree}_{\rm cont}(x) &= \gamma_\rho \Gamma^F \gamma_\sigma \; \frac{1}{8
  \pi^2} \left[ 2\frac{x_\rho x_\sigma}{(x^2)^2}
- \frac{ \delta_{\rho \sigma} }{x^2} \right] \,.
\end{split}
\ee

Let us first calculate the gauge term in the propagator. We have argued that
this will be 
\begin{equation}
\ggcf \xi \ {\cal F}\left[ S^{\rm tree}_{\rm cont}(x)
\Big( F_0 + \gamma_E + \log \frac{x^2}{4 a^2} \Big) \right] \,,
\end{equation}
where ${\cal F}$ is the Fourier transform. Some details for performing ${\cal
F}$ are given at the end of this subsection. From
\begin{equation}
{\cal F}\left[ S^{\rm tree}_{\rm cont}(x) \right]
= \frac{1}{{\rm i} \pslash}
\end{equation}
and 
\begin{equation}
{\cal F}\left[  S^{\rm tree}_{\rm cont}(x) \log \frac{x^2}{4 a^2}  \right] =
\frac{1}{{\rm i} \pslash}[1 -2 \gamma_E -\log(a^2 p^2) ]
\label{Ftransform}
\end{equation}
we get
\begin{equation}
S(p, g^2, \xi) =  S(p, g^2, \xi=0) +
\frac{1}{{\rm i} \pslash}\; \ggcf\, \xi \; [F_0 -\gamma_E +1 -\log(a^2 p^2)] \,,
\label{Sgaugeterm}
\end{equation}
which agrees with eqs.~(\ref{Sigma}), (\ref{gaugedep}) in the body of the
paper. 

Now we calculate the gauge dependent parts of the {\it non-amputated}
three-point functions: 
\begin{equation}
{\cal F} \left[ G^{\rm tree}_{\rm cont}(x)
\left(F_0 +\gamma_E + \log  \frac{x^2}{4 a^2} \right) \right]
= \frac{1}{{\rm i}\pslash}\, \Gamma^X \frac{1}{{\rm i}\pslash}
\left[ F_0 - \gamma_E + 2 -\log(a^2 p^2) \right]
+ \frac{1}{2}\, \frac{\gamma_\rho\, \Gamma^X \gamma_\rho}{p^2} \,,
\rule[-7mm]{0mm}{8mm}
\end{equation}
which leads to
\begin{equation}
G^X(p,g^2,\xi) =  G^X(p,g^2,0)
+ \ggcf \xi \left\{
\frac{1}{{\rm i}\pslash} \Gamma^X \frac{1}{{\rm i}\pslash}
\left[ F_0 - \gamma_E + 2 -\log(a^2 p^2) \right]
+ \frac{\gamma_\rho \Gamma^X \gamma_\rho}{2 p^2}
\right\}\,,\rule[-7mm]{0mm}{8mm} 
\end{equation}
where $\Gamma^X$ is defined in~(\ref{GammaDef}). When we amputate this using
the gauge dependent quark propagator~(\ref{Sgaugeterm}) we find
\begin{equation}
\Lambda^X(p,g^2,\xi) =  \Lambda^X(p,g^2,0)
+ \ggcf \xi \left\{ \Gamma^X \left[-F_0 + \gamma_E +\log(a^2 p^2) \right]
- \frac{1}{2 p^2}
\pslash \gamma_\rho \Gamma^X \gamma_\rho \pslash \right\}\,,
\rule[-7mm]{0mm}{8mm} 
\end{equation}
valid for general $\Gamma^X$.
The final term simplifies when we put in specific choices for $\Gamma^X$,
giving
\begin{eqnarray}
\Lambda^{S,P}(p,g^2,\xi)\!\!\! &=& \!\!\!  \Lambda^{S,P}(p,g^2,0)
+\ggcf \xi \left[ -F_0 + \gamma_E -2 +\log(a^2 p^2) \right]
\{\bbbone,\gamma_5\} 
\,,
\nonumber\\[0.5em]
\Lambda^{V,A}_\mu(p,g^2,\xi)\!\!\! &=& \!\!\!  \Lambda^{V,A}_\mu(p,g^2,0)
+ \ggcf \xi \left\{ \gamma_\mu
\left[-F_0 + \gamma_E -1 +\log(a^2 p^2) \right]
+ 2 \frac{p_\mu\! \pslash}{p^2} \right\}  \{\bbbone,\gamma_5\}
\,,
\nonumber\\[0.5em]
\Lambda^T_{\mu \nu}(p,g^2,\xi)\!\!\! &=& \!\!\!  \Lambda^T_{\mu \nu}(p,g^2,0)
+ \ggcf \xi \left[  -F_0 + \gamma_E +\log(a^2 p^2)
\right] \sigma_{\mu \nu} \gamma_5 \,,
\end{eqnarray}
which reproduces all the gauge dependent terms in eq.~(\ref{3PF}).
Equation~(\ref{exactpos}) still holds if a quark mass is included,
so it should also reproduce the mass dependence of the gauge terms
previously calculated in~\cite{Capitani:2000xi}.

The integrals needed to perform the Fourier transform into momentum space are
easily found by introducing an additional integration over $\alpha$, to make
the $x$ integrations into Gaussians. As an example, let us transform $S^{\rm
  tree}_{\rm cont}(x) ({x^2}/{4 a^2})^\delta$ into momentum space:
\begin{eqnarray}
\frac{1}{2 \pi^2} \gamma^\mu
\int d^4 x \;   e^{-{\rm i} p x}  \frac{ x^\mu}{(x^2)^2}
\left(  \frac {x^2}{4 a^2} \right)^\delta &=&
\frac{1}{2 \pi^2} \gamma^\mu \int d^4 x \; e^{-{\rm i} p x} x^\mu
\int_0^\infty \frac{d \alpha \; \alpha^{1-\delta}}
{\Gamma(2-\delta)} e^{-\alpha\, x^2} (4 a^2)^{-\delta}
\nonumber \\[0.5em]
&=& -\frac{{\rm i}\!\pslash}{4}\, \frac{ (4 a^2)^{-\delta}}{\Gamma(2-\delta)}
\int_0^\infty d\alpha \; \alpha^{-2-\delta} e^{-{p^2}/{4 \alpha} } \\[0.5em]
&=& -\frac{{\rm i}\! \pslash}{p^2} \,
\frac{\Gamma(1 + \delta) }{\Gamma(2-\delta)}\, (a^2 p^2)^{-\delta}
\,. \nonumber 
\end{eqnarray}
Expanding both sides to first order in $\delta$, we get the
result~(\ref{Ftransform}).
The integrals needed for the Fourier transform of the three-point functions
can be obtained in the same way as outlined above. We obtain
\begin{equation}
\int d^4 x \; e^{-{\rm i} p x} \, \frac{1} {x^2}
\left( 1 +\delta \log \frac{x^2}{4 a^2} \right)
= \frac{4 \pi^2}{p^2}
\left\{ 1 +\delta \left[ -2\gamma_E -\log(a^2 p^2) \right]\right\}
\end{equation}
and
\begin{equation}
\begin{split}
\int d^4 x \,  e^{-{\rm i} p x} \, \frac{x_\rho x_\sigma}{(x^2)^2}
\left( 1 +\delta \log \frac{x^2}{4 a^2} \right)
&= 2 \pi^2 \frac{\delta_{\rho \sigma} }{p^2}
\left\{ 1 +\delta \left[1 -2\gamma_E -\log(a^2 p^2) \right] \right\} \\
&\phantom{=\;} - 4 \pi^2  \frac{p_\rho p_\sigma}{(p^2)^2}
\left\{ 1 +\delta \left[2 -2\gamma_E -\log(a^2 p^2) \right]\right\} \,.
\end{split}
\end{equation}

\noindent
{\bf Gauge Dependence in Momentum Space}

Here we calculate explicitly the gauge dependent one-loop contributions
to the quark self-energy and the amputated Green functions using the Feynman
rules. We now set $a = 1$ again.

\noindent
{\it The generalised lattice Ward identity}

As can be easily checked, the generalised lattice Ward identity for Wilson and
overlap fermions to lowest order is of the form
\be
S_{W,N}^{-1}(p_2)-S_{W,N}^{-1}(p_1)= \sum_\mu {\widehat {(p_1-p_2)}}_\mu
V_{1\,\mu}^{W,N}(p_2,p_1) 
= {\widehat {(p_1-p_2)}} V_{1}^{W,N}(p_2,p_1)\,,
\ee
where inverse propagators and vertices are defined in Appendix A.
>From this identity it can be seen that
\be
\begin{split}
\bbbone &= -\hat k\,  V_{1}^{W,N}(p,p+k) \, S_{W,N} (p+k) + S_{W,N}^{-1}(p)
S_{W,N}(p+k) \\ 
&= -S_{W,N} (p+k) \, \hat k\,  V_{1}^{W,N}(p+k,p) + S_{W,N}(p+k)
S_{W,N}^{-1}(p) \,,
\label{identity}
\end{split}
\ee
which leads in the limit $p\to 0$ to
\be
\bbbone = -\hat k\,  V_1^{W,N}(0,k) \, S_{W,N} (k)=- S_{W,N} (k) \, \hat k\,
V_1^{W,N}(k,0) \,. 
\label{trafo1}
\ee
In addition we have the ordinary lattice Ward identity to lowest order
(with zero gauge boson momentum)
\be
\frac{\partial}{\partial p_\mu} S_{W,N}^{-1}(p)= - V_{1\,\mu}^{W,N}(p,p) \,.
\label{derivs}
\ee
Differentiating (\ref{identity}) with respect to $p_\mu$, and taking the limit
$p \to 0$, we get 
\be
\frac{\partial}{\partial p_\mu} S_{W,N}(p+k)\Bigr|_{p=0} \,
\hat k\, V_1^{W,N}(k,0)=
- S_{W,N}(k)\frac{\partial}{\partial p_\mu} \, \hat k\,
V_1^{W,N}(p+k,p)\Bigr|_{p=0}\,  +{\rm i} S_{W,N}(k) \, \gamma_\mu \,.
\label{trafo2}
\ee
Using $S_{W,N}^{-1} S_{W,N}=1$ and (\ref{derivs}), (\ref{trafo2}), another
useful form can be derived:
\be
\frac{\partial} {\partial p_\mu} \, \hat k\, V_1^{W,N}(p+k,p) \Bigr|_{p=0}=
V_{1\,\mu}^{W,N}(k,k) + {\rm i} \gamma_\mu\,.
\label{trafo3}
\ee

\noindent
{\it Gauge dependent one-loop corrections}

We consider the gauge dependent part of the gluon propagator in a general
covariant gauge, 
\be
D_{\mu\nu}^{\rm gauge}(k)= - \frac{\hat k_\mu \hat k_\nu}{ ({\hat k^2})^2 }\,,
\label{gluonpropgauge}
\ee
and use the short-handed notation for the D-dimensional one-loop integration
variable 
\be
\int_k\equiv \int \frac{d^{\rm D} k}{(2 \pi)^{\rm D}}  \,.
\ee
For finite integrals we replace the dimension D by four.
The basic divergent lattice integral (in dimensional regularisation) is
\be
\int_k \frac{1}{ ({\hat k^2})^2 }
=\frac{1}{16 \pi^2} \left[ \frac{2}{{\rm D}-4} + F_0 - \log 4 \pi -\log
  (a^2\mu^2 ) \right]\,. 
\ee
Furthermore, we use the finite lattice integral
\be
\int_k \frac{1}{ {\hat k^2} }
=Z_0
\,.
\ee
First we demonstrate that the gauge dependent contribution of one-loop
corrections to local quark bilinears does not depend on the particular
representation (Wilson or overlap) of the lattice fermions.
Since the local operators do not contain gluon operators, we have to consider
only the vertex correction to the amputated Green function:
\be
I^X(p)=g^2 C_F \sum_{\mu\nu}
\int_k
V_{1\,\mu}^{W,N}(p,p+k)\, S_{W,N}(p+k) \,{\Gamma}^X\, S_{W,N}(p+k)\,
V_{1\,\nu}^{W,N}(p+k,k) \, D_{\mu\nu}^{\rm gauge}(k)\,. 
\ee
The corresponding correction is ultraviolet (UV) logarithmically divergent.
Using the technique for analytic handling of the divergences, we have to
consider 
\be
I^{X\,{\rm lat}}(p) = I^X(0) + I^{X\,{\rm cont}}(p)\,.
\ee
With the propagator~(\ref{gluonpropgauge}), and using eq.~ (\ref{trafo1}), we
get at zero momentum the expected result
\be
\begin{split}
I^X(0) &= - g^2 C_F
\int_k \frac{1}{ ({\hat k^2})^2 }
\hat k\, V_{1}^{W,N}(0,k)\, S_{W,N}(k) \,{\Gamma}^X\, S_{W,N}(k)\, \hat k\,
V_{1}^{W,N}(k,0) \\[0.7ex] 
&=-g^2 C_F\,{\Gamma}^X \int_k \frac{1}{ ({\hat k^2})^2 } \,.
\end{split}
\ee
The pole cancels with that of the continuum contribution, $I^{X\,{\rm
cont}}(p)$, while the one-loop correction is independent of the form of the
lattice propagator as result of the Ward identity.

For operators with derivatives one has to consider the Taylor expansion of the
lattice integrals up to the corresponding order of the UV divergence.
In that case additionally the $O(g)$ and $O(g^2)$ contributions of the
operators have to be considered. It is not difficult to perform a similar
proof. 

The sunrise diagram (i.e. the left diagram of Fig.~\ref{fig1}) for the quark
self-energy is of the form
\be
I^{\rm sunrise}(p)= g^2 C_F \sum_{\mu\nu}
\int_k V_{1\,\mu}^{W,N}(p,p+k) S_{W,N}(p+k) V_{1\,\nu}^{W,N}(p+k,p)
D_{\mu\nu}^{\rm gauge}(k)\,. 
\label{sunrise}
\ee
Since that one-loop integral is UV linearly divergent, we have to calculate
(using dimensional regularisation)
\bea
I^{\rm sunrise}(p)= I^{\rm sunrise}(0) + \sum_\alpha p_\alpha
\frac{\partial}{\partial p_\alpha} 
I^{\rm sunrise}(p)\Bigr|_{p=0} + I^{\rm cont} (p) \,.
\eea
For the finite tadpole contribution (D=4) we have
\be
I^{\rm tadpole}(p)= g^2 C_F \sum_{\mu\nu}\int_k V_{2\,\mu\nu}^{W,N}(p,p,k,-k)
D_{\mu\nu}^{\rm gauge}(k)\,. 
\label{tadpole}
\ee
The gauge dependent quark self-energy contribution $I^{\rm lat}(p)$
is then given as sum of (\ref{sunrise}) and (\ref{tadpole}).
To show the independence on the lattice fermion representation, we transform
the difference
$I^{\rm lat}(p)-I^{\rm cont}(p)$, using eqs.~(\ref{trafo1})-(\ref{trafo3}),
to the form
\bea
I^{\rm lat}(p) - I^{\rm cont}(p) =
g^2 C_F \left[{{\rm i} \pslash \int_k \frac{1}{ ({\hat k^2})^2}
    -(J_1^{W,N}+J_2^{W,N}+J_3^{W,N} ) } \right] 
\eea
where
\bea
\label{J1}
J_1^{W,N}(\bbbone)&=&
-\int_k \frac{1}{ ({\hat k^2})^2 }
\hat k\, V_1^{W,N}(k,0)\,,
\\
\label{J2}
J_2^{W,N}(\pslash)&=& -\sum_\mu p_\mu \int_k \frac{1}{ ({\hat k^2})^2 }
 \left[V_{1\,\mu}^{W,N}(k,k) + {\rm i} \gamma_\mu \right]\, ,
\\
\label{J3}
J_3^{W,N}(\bbbone,\pslash)&=&
\int_k \frac{1}{ ({\hat k^2})^2 }
\hat k\, V_{2}^{W,N}(p,p,k,-k)\, \hat k  \,.
\eea
It remains to show that the sum of integrals $J_1^{W,N}+J_2^{W,N}+J_3^{W,N}$
vanishes.
For Wilson fermions we immediately obtain
\bea
J_1^W(\bbbone)=\frac{r}{2} Z_0\,,\quad
J_2^W(\pslash)=-\frac{1}{8}\, {\rm i} \pslash \, Z_0\,, \quad
J_3^W(\bbbone,\pslash)=\frac{1}{8}\, {\rm i} \pslash \, Z_0 - \frac{r}{2}
Z_0\,.
\eea
Therefore the sum is zero.

\noindent
{\it Integrals for overlap fermions}

In the following we use the abbreviations (\ref{ob}). Using (\ref{kV1O}), and
taking into account the symmetry in the integration, we get immediately
\be
J_1^N(\bbbone)
= \rho \int_k \frac{1} {({\hat k^2})^2 }    \frac{\omega+ b }{\omega}\,.
\label{sunrise0}
\ee
To calculate $J_2^N$ we use the symmetric part of
$V_{1\,\mu}^N(k,k)$ in $k$ and get
\bea
J_2^{N}(\pslash)=
-{\rm i} \sum_\mu p_\mu \gamma_\mu \int_k\frac{1}{ ({\hat k^2})^2 }
\frac{1}{ \omega^3}
\biggl[
\omega^2(\omega- \rho \cos k_\mu) + \rho \, \sin^2 k_\mu (\cos k_\mu + r b )
\biggr]
\label{J2o}
\,.
\eea
To calculate $J_3^N$, we split it into two pieces, $J_3^N = J_{31}^N+J_{32}^N$,
according to eq.~(\ref{qqggover}). After
symmetrisation and taking $\sum_\mu V_{21\,\mu\mu}^N(p,p)$ 
in the zero lattice spacing limit, we find for the $k$ independent part
\be
J_{31}^N(\pslash)\equiv
\int_k \frac{1}{ ({\hat k^2})^2 } \,
\hat k\, V_{21}^N(p,p)\, \hat k
= {\rm i} \pslash \, \frac{1}{8} \left( 1 - 4 \frac{r}{\rho} \right) Z_0 \,.
\label{J31}
\ee
For the cancellation to be shown, we write this result in the form
\be
J_{31}^N(\pslash)= {\rm i} \sum_\mu p_\mu \gamma_\mu \int_k\frac{1}{ ({\hat
k^2})^2 } \left( 1 - \cos k_\mu -\frac{b+\rho}{\rho} \right) \,.
\label{J31prime}
\ee
The $k$ dependent vertex part arising from (\ref{v2over2}) contributes both
to the unit matrix $\bbbone$ and to $\pslash$.
After integrating over $k$ we have to take the zero spacing limit.
This is achieved by performing a Taylor expansion around $p=0$
up to linear terms in $p$: 
\be
\begin{split}
J_{32}^N(\bbbone,\pslash)&=
\int_k \frac{1}{ ({\hat k^2})^2 }
\left[\hat k\, V_{22}^N(0,0,k,-k)\,\hat k +\sum_\mu p_\mu
  \frac{\partial}{\partial p_\mu} 
\hat k\, V_{22}^N(p,p,k,-k)\,\hat k\biggr|_{p=0} \right]\\
&= J_{32}^N(\bbbone)+ J_{32}^N(\pslash)\,.
\end{split}
\ee
Using the form~(\ref{kV22knew})
we get
\be
\begin{split}
J_{32}^N(\bbbone)&= -\rho  \int_k
\frac{1}{({\hat{k}^2})^2}\frac{\omega+b}{\omega}\,, \\
J_{32}^N(\pslash)&=
\sum_\mu {\rm i}\, p_\mu \gamma_\mu
\int_k\frac{1}{({\hat k^2})^2 }
\biggl[
\frac{b+\rho}{\rho} +\frac{ (\omega-\rho) \cos k_\mu}{\omega}
+\frac{\rho \sin^2 k_\mu \,(\cos k_\mu + r b)}{\omega^3}
\biggr] \, .
\end{split}
\ee
We see that $J_{32}^N(\bbbone)$ cancels the contribution from the sunrise
diagram~(\ref{sunrise0}), 
and
\be
J_2^N(\pslash)+ J_{31}^N(\pslash)+J_{32}^N(\pslash) = 0 \, .
\ee

\section{Appendix C}

   In this appendix we give a more complete derivation 
 of the mean field overlap fermion
 propagator and tadpole improved $\rho$, 
 without having to expand in powers of $a$. 

  To calculate the mean field overlap fermion propagator, we
 start from the
 mean field value of the Wilson-Dirac operator in momentum space,
 \begin{equation}
 D_W^{\rm MF}(u_0,r)  = \frac{1}{a} \left(
  u_0  \sum_\mu {\rm i}\, \gamma_\mu \sin a p_\mu + 4 r
 - u_0\,  r \sum_\mu \cos a p_\mu \right) \,. 
  \end{equation}
    This can be written in terms of the tree-level Dirac operator 
 \begin{equation} 
  D_W^{\rm MF}(u_0,r) = u_0 D_W^{\rm tree}(r) + \frac{4r}{a}(1 -u_0)\,. 
 \end{equation}  
 Let us now calculate the overlap operator from $D_W^{\rm MF}$. 
 The first step is to find the $X$ operator
 \begin{equation}
\begin{split} 
 X^{\rm MF}(u_0,r,\rho) &\equiv  D_W^{\rm MF}(u_0,r) -\frac{\rho}{a} \\
 &= u_0 \left[ D_W^{\rm tree}(r) - \frac{ \rho -4 r(1-u_0) }{a\, u_0}
  \right] \\
 &= u_0 X^{\rm tree}(r, \rho^{TI})\,,  
\end{split} 
\end{equation} 
 where
 \begin{equation}
  \rho^{TI} = \frac{ \rho -4 r(1-u_0) }{u_0} \,. 
 \end{equation} 
 We are now ready to calculate the mean field overlap operator 
 \begin{equation}
\begin{split}
 D_N^{\rm MF}(u_0,r,\rho) &= \frac{\rho}{a}  \left[ 1 + 
  \frac{ X^{\rm MF}(u_0,r,\rho) }{ \sqrt{ {X^{\rm MF}}^\dagger(u_0,r,\rho)
   X^{\rm MF}(u_0,r,\rho) }} \right] \\ 
 &= \frac{\rho}{a} \left[ 1 +
  \frac{  X^{\rm tree}(r, \rho^{TI} ) }
  { \sqrt{ {X^{\rm tree}}^\dagger(r,\rho^{TI}) X^{\rm tree}(r,\rho^{TI}) }}
  \right] \\ 
 &=  \frac{\rho}{\rho^{TI}} D_N^{\rm tree}(r,\rho^{TI})
 \equiv  Z_\psi^{\rm MF}  D_N^{\rm tree}(r,\rho^{TI}) \,.
\end{split}  
\end{equation}
 In other words, the mean field overlap operator is proportional to 
 the tree-level overlap operator, calculated with the same $r$ but with
 a tadpole improved $\rho$. The constant of proportionality is
 \begin{equation}
 Z_\psi^{\rm MF} = \frac{\rho}{\rho^{TI} } = 
  \frac{ \rho u_0} { \rho -4 r(1-u_0) } \,. 
 \end{equation}

\end{document}